\documentclass[
  reprint,
  superscriptaddress,
  amsmath,amssymb,
  aps,
  pra,
  floatfix,
  twocolumn
]{revtex4-2}

\usepackage{graphicx}
\usepackage{dcolumn}
\usepackage{bm}
\usepackage{amssymb}
\usepackage{physics}
\usepackage{mathtools}
\usepackage{tikz}
\usepackage{amsmath}
\usepackage{comment}

\DeclareMathOperator*{\argmax}{argmax}

\usepackage[
  colorlinks=true,
  allcolors=blue,
  breaklinks=true,
  pdfpagemode=UseNone,
  pdfstartview=FitH,
  bookmarksnumbered=true,
  bookmarksopen=false,
]{hyperref}

\begin{document}

\preprint{APS/123-QED}

\title{Group-Theoretic Reinforcement Learning of Dynamical Decoupling Sequences}

\author{Charles Marrder}
\affiliation{
JILA, Department of Physics, University of Colorado Boulder, Boulder, Colorado 80309, USA
}
\author{Shuo Sun}%
\affiliation{
JILA, Department of Physics, University of Colorado Boulder, Boulder, Colorado 80309, USA
}
\author{Murray J. Holland}%
\affiliation{
JILA, Department of Physics, University of Colorado Boulder, Boulder, Colorado 80309, USA
}

\date{December 10, 2025}

\begin{abstract}
Dynamical decoupling seeks to mitigate phase decoherence in qubits by applying a carefully designed sequence of effectively instantaneous electromagnetic pulses.
Although analytic solutions exist for pulse timings that are optimal under specific noise regimes, identifying the optimal timings for a realistic noise spectrum remains challenging.
We propose a reinforcement learning (RL)-based method for designing pulse sequences on qubits. 
Our novel action set enables the RL agent to efficiently navigate this inherently non-convex optimization landscape.
The action set, derived from Thompson's group $F$, is applicable to a broad class of sequential decision problems whose states can be represented as bounded sequences.
We demonstrate that our RL agent can learn pulse sequences that minimize dephasing without requiring explicit knowledge of the underlying noise spectrum.
This work opens the possibility for real-time learning of optimal dynamical decoupling sequences on qubits which are dephasing-limited.
The model-free nature of our algorithm suggests that the agent may ultimately learn optimal pulse sequences even in the presence of unmodeled physical effects, such as pulse errors or non-Gaussian noise.
\end{abstract}

\maketitle

\section{Introduction}
\label{sec:intro} 

Dynamical decoupling (DD) is a Hamiltonian-engineering technique that applies carefully designed sequences of control pulses to a quantum system in order to modify its intrinsic dynamics.
Originally inspired by nuclear magnetic resonance methods, DD was developed to suppress decoherence by dynamically decoupling an open quantum system from its environment \cite{Viola_DynamicalSuppressionDecoherence_1998, Viola_DynamicalDecouplingOpen_1999}.
Over the past two decades, DD has proven especially valuable in quantum information processing, enhancing the performance of quantum algorithms \cite{Jurcevic_DemonstrationQuantumVolume_2021, Pokharel_DemonstrationAlgorithmicQuantum_2023} and quantum error correction protocols \cite{Chen_ExponentialSuppressionBit_2021, Krinner_RealizingRepeatedQuantum_2022, Sundaresan_DemonstratingMultiroundSubsystem_2023}.
Crucially, DD is regarded as a practical means of reducing physical error rates to levels where quantum error correction can be implemented with reasonable resource overheads \cite{Ng_CombiningDynamicalDecoupling_2011, Paz-Silva_OptimallyCombiningDynamical_2013}.

The field of dynamical decoupling (DD) has expanded significantly since its inception, with applications now spanning noise spectroscopy \cite{Degen_QuantumSensing_2017, Vezvaee_FourierTransformNoise_2024}, quantum metrology \cite{Ji_DynamicalDecouplingassistedQuantum_2021, Kim_SuppressionSpinbathLowfrequency_2025}, and increasingly sophisticated Hamiltonian-engineering schemes \cite{Conrad_TwirlingHamiltonianEngineering_2021, Tyler_HigherorderMethodsHamiltonian_2023, Zhou_RobustHamiltonianEngineering_2024, Zhao_HigherorderProtectionQuantum_2025}.
Error suppression via DD remains an active area of research, as evidenced by the development of new performance bounds \cite{Hahn_EfficiencyDynamicalDecoupling_2025, Burgelman_LimitationsDynamicalError_2025}, studies of DD in qudits \cite{Yuan_PreservingMultilevelQuantum_2022, Tripathi_QuditDynamicalDecoupling_2025}, and extensions to multi-qubit systems \cite{Tripathi_SuppressionCrosstalkSuperconducting_2022}.
More recently, there has been growing interest in constructing novel DD sequences using optimization and machine learning techniques \cite{Li_DesigningArbitrarySingleaxis_2021, Cai_OptimizingContinuousDynamical_2022, Rahman_LearningHowDynamically_2024, Tong_EmpiricalLearningDynamical_2025}.
Building on this progress, the present work demonstrates that discrete reinforcement learning provides an effective framework for optimizing DD sequences. 

The prototypical example of DD is the spin echo~\cite{Hahn_SpinEchoes_1950}, which refocuses qubit coherence under inhomogeneous dephasing noise using a single $\pi$ pulse.
The Carr–Purcell–Meiboom–Gill (CPMG) sequence \cite{Carr_EffectsDiffusionFree_1954, Meiboom_ModifiedSpinEchoMethod_1958} extends this idea by repeating the spin echo periodically.
Although these techniques were originally developed to mitigate the dephasing of nuclear spin states in nuclear magnetic resonance experiments, DD has since been shown to suppress noise in a wide range of quantum information systems \cite{Suter_ColloquiumProtectingQuantum_2016}.
Several extensions of CPMG have been proposed to enhance performance under different noise spectra.
For example, the Uhrig dynamical decoupling (UDD) sequence is analytically optimal for noise with a hard frequency cutoff, whereas CPMG performs better for soft cutoff spectra \cite{Uhrig_ExactResultsDynamical_2008}.
Other well-known protocols include periodic dynamical decoupling (PDD) and concatenated dynamical decoupling (CDD).
In PDD, pulses are applied at uniformly spaced intervals, while in CDD, higher-order sequences are generated recursively by concatenating lower-order ones \cite{Khodjasteh_FaultTolerantQuantumDynamical_2005, Khodjasteh_PerformanceDeterministicDynamical_2007}.
In general, the specific timing structure of a DD sequence determines its ability to suppress different types of noise.

When evaluating the performance of traditional DD pulse sequences, one typically assumes a specific model for the system–bath interactions and the corresponding noise spectrum \cite{Suter_ColloquiumProtectingQuantum_2016}.
In practice, however, the noise spectrum is often unknown and may not be easily categorized as having either a hard or smooth cutoff.
One way to solve this issue is to reconstruct the noise spectrum and optimize pulse timings accordingly, and there has been great progress in spectral reconstruction protocols over the last two decades \cite{Alvarez_MeasuringSpectrumColored_2011, Bylander_NoiseSpectroscopyDynamical_2011, Yuge_MeasurementNoiseSpectrum_2011, Norris_QubitNoiseSpectroscopy_2016, Paz-Silva_MultiqubitSpectroscopyGaussian_2017, Szankowski_EnvironmentalNoiseSpectroscopy_2017, Szankowski_AccuracyDynamicaldecouplingbasedSpectroscopy_2018, Paz-Silva_ExtendingCombbasedSpectral_2019, Sung_NonGaussianNoiseSpectroscopy_2019, VonLupke_TwoQubitSpectroscopySpatiotemporally_2020, Chalermpusitarak_FrameBasedFilterFunctionFormalism_2021, Dong_ResourceefficientDigitalCharacterization_2023, McCourt_LearningNoiseDynamical_2023, Khan_MultiaxisQuantumNoise_2024, Vezvaee_FourierTransformNoise_2024, Wang_DigitalNoiseSpectroscopy_2024, Dong_EfficientLearningOptimizing_2025, Wang_BroadbandSpectroscopyQuantum_2025}.
However, these protocols only approximate the true underlying physical dynamics by making assumptions about the structure of the environmental noise (classical vs.\ quantum, Gaussian vs.\ non-Gaussian, pure dephasing vs.\ multi-axis), the characteristics of the control sequences (perfect vs.\ noisy, single-axis vs.\ multi-axis), or the noisiness of state preparation and measurement (SPAM). 
Moreover, most protocols do not account for non-stationary noise, so if the noise spectrum drifts over time, the reconstruction procedure generally must be repeated, thus demanding resource efficiency.
There is so far no resource-efficient quantum noise spectroscopy protocol capable of characterizing potentially quantum, non-Gaussian, non-Markovian, multi-axis noise with noisy SPAM and noisy, multi-axis controls.

These challenges motivate the development of a general-purpose learning-based framework capable of rapidly discovering non-analytic DD pulse sequences that are optimal for an unknown, potentially time-varying noise environment.
Ideally, such an approach would also overcome the limitations of traditional analytical and numerical optimization techniques. This work represents a step toward this goal.

Although various optimization protocols could, in principle, be used to address this problem, our goal is to employ a method that can identify an optimal solution with a minimal number of objective function evaluations.
This constraint is particularly important for experimental implementation, where the objective function depends on measurements of quantum state fidelity.
Depending on the experimental platform, these measurements can be time-consuming and resource-intensive.
Furthermore, because the underlying noise spectrum of a quantum system may drift over time, it is desirable to learn the optimal pulse sequence as rapidly as possible.
We therefore seek an optimization strategy that is both sample-efficient and robust to the nonconvex nature of the pulse sequence optimization landscape, ensuring reliable convergence even in the presence of local optima.

These constraints make traditional optimization protocols less suitable for our purposes.
Gradient-based methods, in particular, tend to require a prohibitive number of objective function evaluations.
For instance, optimizing a pulse sequence consisting of $N$ pulses using finite-difference gradient estimation would require at least $2N$ evaluations per iteration.
This cost could, in principle, be reduced if a model for the gradient of the fidelity with respect to the $N$ pulse times were available.
However, constructing such a model would require detailed knowledge of the decoherence process, information that is not known {\em a priori}, and any resulting model would likely only approximate the true physical dynamics.

Moreover, experimental implementations of DD are noisy, and the objective function must be estimated from fidelity measurements.
Each fidelity estimate carries inherent statistical uncertainty, further compounded by shot-to-shot pulse errors and the stochastic nature of open quantum system dynamics. Stochastic optimization methods such as simulated annealing and evolutionary algorithms can be used to optimize stochastic objective functions, and they are often robust to local minima. However, these methods do not learn, completely ignoring information about past trajectories through the search space which could potentially enable more efficient searching \cite{Sutton_ReinforcementLearningIntroduction_2018}.

Considering these constraints, we propose that off-policy, model-free reinforcement learning with discrete actions is an especially suitable approach for this problem.
Algorithms in this class are capable of solving Markov decision processes \cite{Sutton_ReinforcementLearningIntroduction_2018}, and we demonstrate that experimental platforms subject to classical noise can benefit from their use.
Specifically, we show that the Double Deep Q-Network (DDQN) algorithm \cite{vanHasselt_DeepReinforcementLearning_2016}, an algorithm in this class, can successfully optimize ideal $\pi$-pulse times on a single qubit when the action set is based on the generators of Thompson's group $F$ \cite{Cannon_IntroductoryNotesRichard_1996, Belk_ThompsonsGroup_2004, Fordham_MinimalLengthElements_2003} and when
the system dynamics is described by the standard filter function formalism \cite{Uhrig_KeepingQuantumBit_2007, Uhrig_ExactResultsDynamical_2008,Cywinski_HowEnhanceDephasing_2008, Biercuk_DynamicalDecouplingSequence_2011}.
To speed up training and hyperparameter tuning, we simplified the qubit dynamics by assuming classical noise and ideal, instantaneous $\pi$-pulses. We ultimately conducted a study on the agent's performance when starting in different initial sequences, and the results serve as a proof-of-concept for our novel action set and agent design.
Later, we discuss future extensions of our method to scenarios with realistic pulse errors and experimental platforms subject to non-classical noise, providing arguments for why this approach is expected to remain effective under these more complex conditions.

\section{Qubit Dephasing Environment}
\label{sec:physical_model}

We now describe the quantum dynamics underlying our simulations. Specifically, we consider a pure dephasing model, in which the qubit relaxation time $T_1$ is much longer than either the spin-echo dephasing time $T_2$ or the free-induction decay dephasing time $T_2^*$.
Following the convention of \cite{Cywinski_HowEnhanceDephasing_2008}, the qubit’s natural frequency $\Omega$ is subject to classical, wide-sense stationary, Gaussian noise $\beta(t)$ with zero mean, which induces dephasing.
This dynamics is described by the time-dependent Hamiltonian
\begin{equation}
    \hat{H}(t) = \frac{\hbar}{2} \bigl( \Omega + \beta(t) \bigr) \hat{\sigma}_z\,,
\end{equation}
where $\sigma_z$ is the Pauli-Z operator.

To evaluate the performance of a DD sequence, we first initialize the qubit in the $\ket{0}$
state.
At $t=0$, we apply a perfect $\hat{R}_y(\pi/2)$ rotation to prepare the qubit in the $\ket{+}$ state. States on the equator of the Bloch sphere experience the most dephasing, so preparing the qubit in $\ket{+}$ will give us a worst-case estimate of noise suppression. Next, we intersperse free evolution under $\hat{H}(t)$ with a sequence of $\hat{R}_x(\pi)$ pulses. 
We assume these pulses are ideal (meaning they have no errors) and delta-shaped (“bang-bang”) pulses (meaning their dynamics are much faster than those of $\hat{H}(t)$).
The effectiveness of noise suppression depends on the sequence of pulse times $(t_1, \ldots, t_N)$ at which the $N$ pulses are applied. At the total evolution time $T$, we apply a final $\hat{R}_y(-\pi/2)$ rotation and measure the probability of finding the qubit in the $\ket{0}$ state.

As shown in \cite{Bauch_DecoherenceEnsemblesNitrogenvacancy_2020}, the ensemble-averaged probability of measuring $\ket{0}$
at the end of the protocol (equivalently, the fidelity with $\ket{0}$) is
\begin{equation}
    p_\mathrm{avg} = \frac{1}{2} \left( 1 +  e^{-\chi}\right) ,
    \label{eq:avg_prob_of_zero}
\end{equation}
where the attenuation function $\chi$ is defined as
\begin{equation}
    \chi = \frac{1}{2\pi} \int_{0}^{\infty} \mathrm{Re}[S(\omega)] F(\omega, t_1, \ldots, t_N, T) \, d\omega .
    \label{eq:chi}
\end{equation}
Assuming wide-sense stationary, classical noise, the noise spectrum $S(\omega)$ is given by the Fourier transform of the two-time correlation function:
\begin{equation}
    S(\omega) = \int_{-\infty}^\infty e^{i\omega t} \langle \beta(t) \beta(0) \rangle \, dt .
    \label{eq:somega}
\end{equation}
The filter function $F(\omega, t_1, \ldots, t_N, T)$ is defined as
\begin{equation}
    F(\omega, t_1, \ldots, t_N, T) = \left| \int_{-\infty}^{\infty} e^{i\omega t} f(t, t_1, \ldots, t_N, T) \, dt \right|^2  .
    \label{eq:filter}
\end{equation}
Here, the switching function $f(t, t_1, \ldots, t_N, T)$ alternates between $+1$ and $-1$ each time a pulse is applied. 
The pulse times are constrained to the interval $[0,T]$, and $f(t)$ is zero outside this interval. 
Our notation differs slightly from that of \cite{Cywinski_HowEnhanceDephasing_2008}, where the authors include an additional factor of $\omega^2/2$ in the definition of the filter function.
A detailed derivation of these results is provided in Appendix \ref{app:filter_function_formalism}.

\begin{figure*}[t]
    \centering
    \includegraphics[width=2\columnwidth]{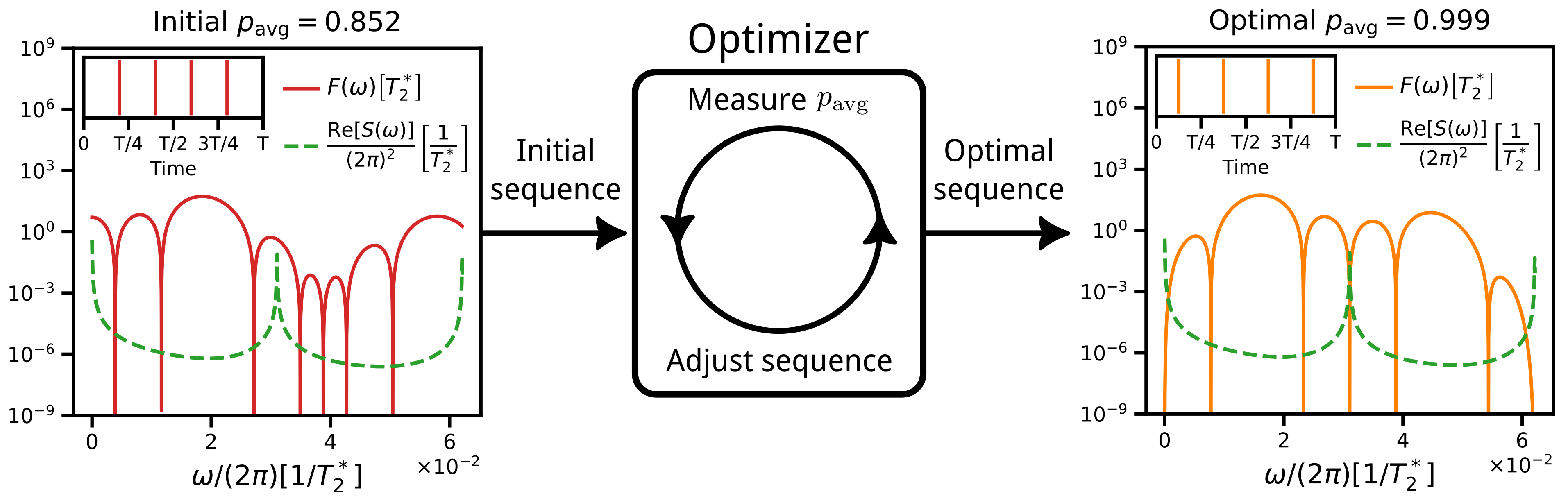}
       \caption{Schematic of our optimization problem. Given some unknown noise spectrum $S(\omega)$ and an initial pulse sequence of fixed length $N$, our goal is to adjust the pulse times to the optimal, unknown sequence using only measurements of $p_\mathrm{avg}$ [Eq.~(\ref{eq:avg_prob_of_zero})]. In the language of the filter function formalism, this corresponds to finding a pulse sequence whose filter function has minimal overlap with the noise spectrum. Within the left panel, the upper left inset illustrates an example initial pulse sequence (red vertical lines), and the rest of this panel shows the corresponding filter function (solid red curve) alongside an example noise spectrum (dotted green curve) composed of a sum of Lorentzian distributions. The middle box represents an optimizer capable of solving this problem and the internal loop defining its algorithm, taking as input the initial pulse sequence timings and outputting the optimal pulse sequence timings. The optimizer will not directly know anything about the noise spectrum or filter, but we illustrate them here to convey the underlying physics problem the optimizer will solve. On the right, we illustrate an example of optimal pulse timings (orange vertical lines) for the given noise spectrum, along with its corresponding filter function (solid orange curve). The filters, noise spectrum, and frequencies are plotted in terms of the free induction decay time $T_2^*$ [Eq.~(\ref{eq:T2star})].}
    \label{fig:optimization_schematic}
\end{figure*}

We now have all the elements necessary to formulate our problem.
Given a fixed noise spectrum $S(\omega)$ and a predetermined number of pulses $N$, our goal is to minimize $\chi$ by designing a filter function $F(\omega, t_1, \ldots, t_N, T)$ that has minimal overlap with $S(\omega)$.
The shape of the filter function can be controlled by adjusting the times at which the $N$ ideal $\hat{R}_x(\pi)$ pulses are applied, but finding these pulse times analytically is challenging. Additionally, the numerical optimization algorithm used to solve this problem should be capable of doing so using only measurements of $p_\mathrm{avg}$, as illustrated in Figure \ref{fig:optimization_schematic}.

Note that in this formulation of the DD problem, we do not allow for the possibility of adding pulses to the sequence.  While adding extra pulses generally improves error suppression in theory, doing so in practice would incur extra timing and resource costs. Also, realistic pulses have some error associated to them, and the error incurred by these extra pulses may outweigh the error suppressed via DD.
Therefore, we consider the case in which an experimental implementation seeks to exploit the most noise suppression possible out of some fixed number of pulses.

\section{Sequential Decision Problem}
\label{sec:decision_problem}

To apply reinforcement learning to this optimization problem, we first cast it as a sequential decision problem.
We start with an initial pulse sequence $s_0 = (t_1, \ldots, t_N)$, where $N$ is the number of pulses to optimize. Our goal is to learn a sequence of actions that transforms $s_0$ into an optimal sequence $s_*$ that minimizes Eq.~(\ref{eq:chi}) after a given number of steps. This problem can be naturally framed as a deterministic Markov decision process, where state transitions and reward payouts deterministically depend only on the current state and the chosen action.

We define the finite set of actions $\mathcal{A}$ as functions that, when applied to a sequence of pulse times, produce a new sequence by transforming the entire interval $[0,T]$. After $M$ steps, the resulting pulse sequence $s_M$ is given by the composition of these actions:
\begin{equation}
    s_M = a_{M-1} \circ a_{M-2} \circ \ldots \circ a_1 \circ  a_0 (s_0) ,
\end{equation}
where $a_j \in \mathcal{A}$ is the action chosen at step $j$. We then apply the final sequence $s_M$, measure the fidelity [Eq.~(\ref{eq:avg_prob_of_zero})], and reset the qubit to the $\ket{0}$ state before starting a new sequence of function compositions. 
We define an episode as the process of beginning with an initial sequence $s_0$, applying $M$ actions, measuring the fidelity, and resetting the qubit.
Our ultimate goal is to construct $s_M = s_*$ using as few fidelity measurements, and thus as few episodes, as possible.

In general, there exists an unknown, continuous and invertible function \begin{equation}
   u:s_0 \mapsto s_* 
\end{equation}
that maps the initial pulse times to the optimal pulse times while preserving the order of the sequence. Technically, $u$ represents an order-preserving homeomorphism on the interval $[0, T]$.
Note that we could instead represent the pulse times as a vector in $[0, T]^N$ and remove the order-preserving constraint on $u$. However, doing so would eliminate the redundancy inherent in pulse time space, arising from the fact that pulses are applied sequentially in time. Exploiting this redundancy allows us to reduce the effective size of the search space.
With this perspective, we can reformulate the optimization problem as follows: given a set of action functions $\mathcal{A}$, which sequence of action function compositions sufficiently approximates the optimal order-preserving homeomorphism $u$?

\section{Thompson's Group \textit{F}}
\label{sec:thompsons_group_f}

To ensure that we can approximate $u$, we seek a set of functions that, through repeated composition, can approximate arbitrarily well any order-preserving homeomorphism. Conveniently, such a set exists in the form of Thompson's group~$F$, which is generated by a finite set of elements \cite{Cannon_IntroductoryNotesRichard_1996, Belk_ThompsonsGroup_2004, Fordham_MinimalLengthElements_2003}. Thompson’s group~$F$ is defined as the set of piecewise linear, order-preserving homeomorphisms on the unit interval $[0, 1]$ that satisfy the following~\cite{Belk_ThompsonsGroup_2004}:
\begin{enumerate}
    \item Breakpoints occur at dyadic rationals of the form $(m/2^n, k/2^\ell)$, with $m, n, k, \ell\in \mathbb{N}$.
    \item Each linear segment's slope is a power of~2.
\end{enumerate}
This structure implies $F$ is dense in the space $\mathrm{Homeo}_+([0,1])$ of order-preserving homeomorphisms of the unit interval $[0, 1]$ with the uniform topology \cite{Bieri_GroupsPLhomeomorphismsReal_2016}.

To build intuition as to why this is true, consider $h\in\mathrm{Homeo}_+([0,1])$. Divide $[0, 1]$ into $2^n$ equally sized subintervals, with $n$ chosen arbitrarily large. Evaluate $h$ at the subintervals' boundary points $p_m = m/2^n$, and round each $h(p_m)$ to the nearest allowable dyadic rational $q_k=k/2^n$. 
By choosing $n$ large enough, the rounded points remain strictly increasing ($q_k<q_{k+1}$ for all $k$), and one can construct an $f\in F$ with breakpoints $(p_m, q_k) = (m/2^n, k/2^n)$. Since $h$ is continuous, sufficiently increasing $n$ ensures that $h$ behaves nearly linearly within each subinterval. 
Consequently, the piecewise-linear $f$ can approximate $h$ to within any desired error tolerance $\epsilon>0$. Writing this in terms of the uniform metric, we have $\sup_{y\in[0,1]} |f(y)-h(y)|<\epsilon$.

Because any function in $F$ can be expressed as a finite composition of its generators, we directly use these generators in our set of actions. Doing so allows a reinforcement learning agent to, in principle, construct sequences of action functions that approximate the optimal pulse-time mapping $u$ with arbitrarily high accuracy.
The generators of $F$ are denoted $x_0$ and~$x_1$~\cite{Belk_ThompsonsGroup_2004}. Our action set consists of these and their inverses, $x_0^{-1}$, and~$x_1^{-1}$, plus the identity function $\mathrm{id}_{[0,1]}$. 
While these functions are defined on the unit interval $[0, 1]$, they can be easily rescaled to map the full time-evolution interval $[0, T]$ to itself.
Including the identity function adds flexibility in constructing sequences of function compositions. 
At any step where it is optimal not to modify the pulse times, the identity allows this directly, avoiding the need to apply a generator and then its inverse in a separate step.
With this reformulation of our optimization problem as a sequential decision problem, we can now leverage well-established reinforcement learning techniques to find optimal pulse sequences.

\section{\label{sec:RL}Reinforcement Learning}

Reinforcement learning (RL) differs from other forms of machine learning in that it learns through interaction with an environment rather than from a fixed dataset, similar to how humans learn to perform multi-step tasks in the real world \cite{Sutton_ReinforcementLearningIntroduction_2018}. 
For example, consider training a computer to play chess. At each turn, the computer observes the current board state, chooses a move, and executes it, thereby changing the state of the game. 
By analyzing the outcomes of many games, the agent learns which moves contribute most to winning or losing. 
Over time, through repeated interaction and feedback, the computer can improve its play, eventually surpassing even expert human players.

Formally, a reinforcement learning (RL) problem involves an agent that interacts with an environment \cite{Sutton_ReinforcementLearningIntroduction_2018}. 
The agent has a set of allowable actions, $\mathcal{A}$, each of which can modify the environment's underlying state. 
At the $i$-th step, the environment provides the agent with an observation $s_i$ of the current state and a scalar reward $r_i \in \mathcal{R} \subset \mathbb{R}$ that quantifies the immediate value of that state. 
Importantly, the observation $s_i$ can be only a principal feature and does not necessarily have to represent the full microscopic state of the environment.

\begin{figure}[htb]
    \centering
    \includegraphics[width=\columnwidth]{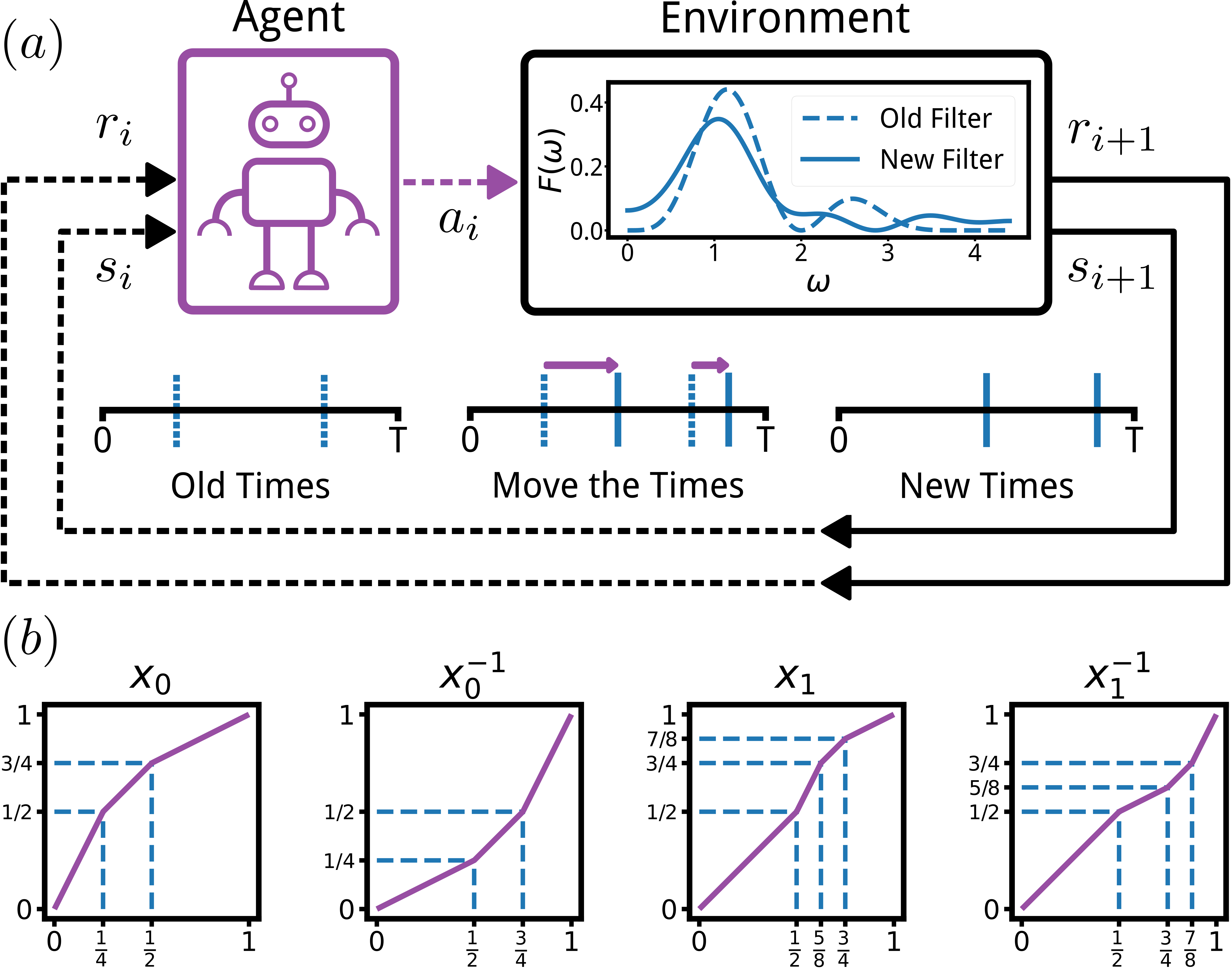}
    \caption{\textbf{(a)} A single reinforcement learning step in our algorithm. At step $i$, the agent observes the current state $s_i$ of the environment and receives a corresponding reward $r_i$. It then selects an action $a_i$ to modify the environment, producing a new state $s_{i+1}$ and reward $r_{i+1}$. In our setting, the state $s_i$ corresponds to the current sequence of pulse times, the reward $r_i$ is defined in Eq.~(\ref{eq:reward}), and actions $a_i$ are functions that transform the interval $[0,T]$, thereby adjusting the pulse times. \textbf{(b)} The generators $x_0$ and $x_1$ of Thompson's group~$F$, along with their inverses. Solid purple lines indicate the function values on the unit interval $[0,1]$, while dotted blue lines mark the breakpoints of each piecewise-linear segment. Evaluated on $[0,T]$ and combined with the identity function $\mathrm{id}_{[0,T]}$, these four functions constitute the action basis for the agent.}
    \label{fig:agent-schematic}
\end{figure}

To select an action $a_i$, the agent consults its behavior policy, which maps a perceived environmental state to a suggested action. 
After executing the action, the agent receives a new observation $s_{i+1}$ and a reward $r_{i+1}$, completing the learning step. Fig.~\ref{fig:agent-schematic}a illustrates a typical learning step for our agent. 
In deterministic environments like ours, the agent’s objective is to maximize the return $G$, or cumulative discounted reward, over many steps,
\begin{equation}
    G=\sum_k \gamma^k r_k ,
\end{equation}
where $\gamma \in [0,1]$ is the discount factor, despite having incomplete knowledge of the environment.

As discussed in Section \ref{sec:intro}, our optimization algorithm must be robust to local minima, sample-efficient, agnostic to the underlying physics, and capable of solving Markov decision processes. These requirements are satisfied by off-policy, model-free, discrete reinforcement learning algorithms, which effectively rules out a large class of alternative RL approaches. The RL algorithm we use is known as the double deep Q-network (DDQN)~\cite{vanHasselt_DeepReinforcementLearning_2016}. While DDQN is capable of solving general Markov decision processes, our optimization problem is fully deterministic. Nevertheless, future extensions of this work may involve stochastic environments. A summary of the mathematics underlying Markov decision processes, Q-learning, and deep Q-learning is provided in Appendix~\ref{app:MathOfQLearning}. The motivation behind our choice of model-free, off-policy, deep RL is given in Appendix \ref{app:our_choice_algo}, while details on our agent’s state encoding, action set, reward function, and hyperparameters can be found in Appendix \ref{app:our_agent_and_environment}.

We set the maximum number of steps per episode to $M = 32$, allowing the agent to construct sufficiently complex homeomorphisms of the time interval $[0,T]$ while remaining able to learn effectively. Choosing $M$ too large would result in overly sparse rewards, whereas a value that is too small would unduly constrain the search space. After training, we select the best pulse sequence discovered by the agent across all episodes

\section{Reinforcement-Learned DD Sequences}
For an unknown noise spectrum, the agent’s goal is to maximize noise suppression using a fixed number $N$ of pulses, relying solely on fidelity measurements. One way to assess whether the agent achieves this goal is to randomly generate a noise spectrum and compare the agent’s optimized sequence to the initial pulse sequence it started from. By repeating this process many times across multiple different initial sequences, we can determine whether the agent consistently improves performance. If it does, this provides strong evidence that our RL approach is effective.

\subsection{Generation of Random Test Spectra}
\label{sec:generation_spectra}

We construct each test spectrum as a sum of real, randomly generated Lorentzian distributions. 
Lorentzian spectra are physically relevant \cite{Klauder_SpectralDiffusionDecay_1962, Bar-Gill_SuppressionSpinbathDynamics_2012, Paladino_1NoiseImplications_2014, Romach_SpectroscopySurfaceInducedNoise_2015, Wudarski_CharacterizingLowFrequencyQubit_2023} and introduce nontrivial structure, allowing us to test the agent on noise spectra for which no analytic solutions exist.
To fairly compare the performance of the initial sequences with the sequences discovered by the agent, we generate each spectrum such that the upper bound on $\chi$ [Eq.~(\ref{eq:chi})] is the same across all noise spectra and for any possible filter function.
Treating $\chi = \int_0^\infty \mathrm{Re}[S(\omega)] F(\omega, t_1,\ldots,t_N,T) \, d\omega/(2\pi)$ as the $L_2$ inner product between $\mathrm{Re}[S(\omega)]$ and $F(\omega, t_1,\ldots,t_N,T)$ over the space of positive frequencies, the Cauchy-Schwarz inequality implies
\begin{eqnarray}
    \chi &\leq& \sqrt{\int_0^\infty \mathrm{Re}[S]^2 \, d\omega / (2\pi) } \sqrt{\int_0^\infty F^2 \, d\omega/(2\pi) } \nonumber\\
    &\leq& \sqrt{\int_0^\infty \mathrm{Re}[S]^2 \, d\omega/(2\pi) } \, ||F||_\mathrm{max}
\end{eqnarray}
where $||F||_\mathrm{max}$ denotes the maximum possible $L_2$ norm of any filter function.

Although $||F||_\mathrm{max}$ is a fixed quantity beyond our control, we ensure that
\begin{equation}
    \sqrt{\int_0^\infty \mathrm{Re}[S(\omega)]^2 \, d\omega / (2\pi)} = ||\mathrm{Re}[S(\omega)]||
\end{equation}
is identical for every randomly generated $S(\omega)$. 
Each Lorentzian in $S(\omega)$ has a full width at half maximum of $2\pi/T$, while the amplitudes are initially drawn randomly from $[0,1]$ and then normalized so that the above quantity equals a chosen constant.
The first Lorentzian is always centered at $\omega=0$ to represent DC noise, and the remaining Lorentzian centers are randomly and uniquely chosen from the set $\{2\pi(j/T) \,| \, j=1, \ldots,2N\}$. 
This ensures that Lorentzians do not overlap excessively, which would make it artificially easier for the agent to minimize the overlap between the noise spectrum and the filter function.

Numerical calculations for randomly generated pulse sequences indicate that the frequency interval $[0, 2\pi(2N/T)]$ contains the vast majority of any filter function's $L_2$ norm over positive frequencies, i.e.,
\begin{equation}
    \frac{\int_0^{2\pi(2N/T)} F^2 \, d\omega/(2\pi)}{\int_0^\infty F^2 \, d\omega/(2\pi)} \gtrapprox 0.99 .
\end{equation}
This further places all possible filter functions on equal footing, since all filters scale as $1/\omega^2$, and Lorentzians centered above $2\pi(2N/T)$  contribute negligibly to $\chi$ compared to those within 
$[0,2\pi(2N/T)]$.

In our simulations, we set $T=1$ and $N=10$, generating 200 random noise spectra, each composed of 5 Lorentzian distributions and normalized so that
\begin{equation}
    \sqrt{\int_0^\infty \mathrm{Re}[S(\omega)]^2 \, d\omega/(2\pi)} = 10.
\end{equation}

\subsection{Agent Performance}

\begin{figure*}[!t]
    \centering
    \includegraphics[width=\textwidth]{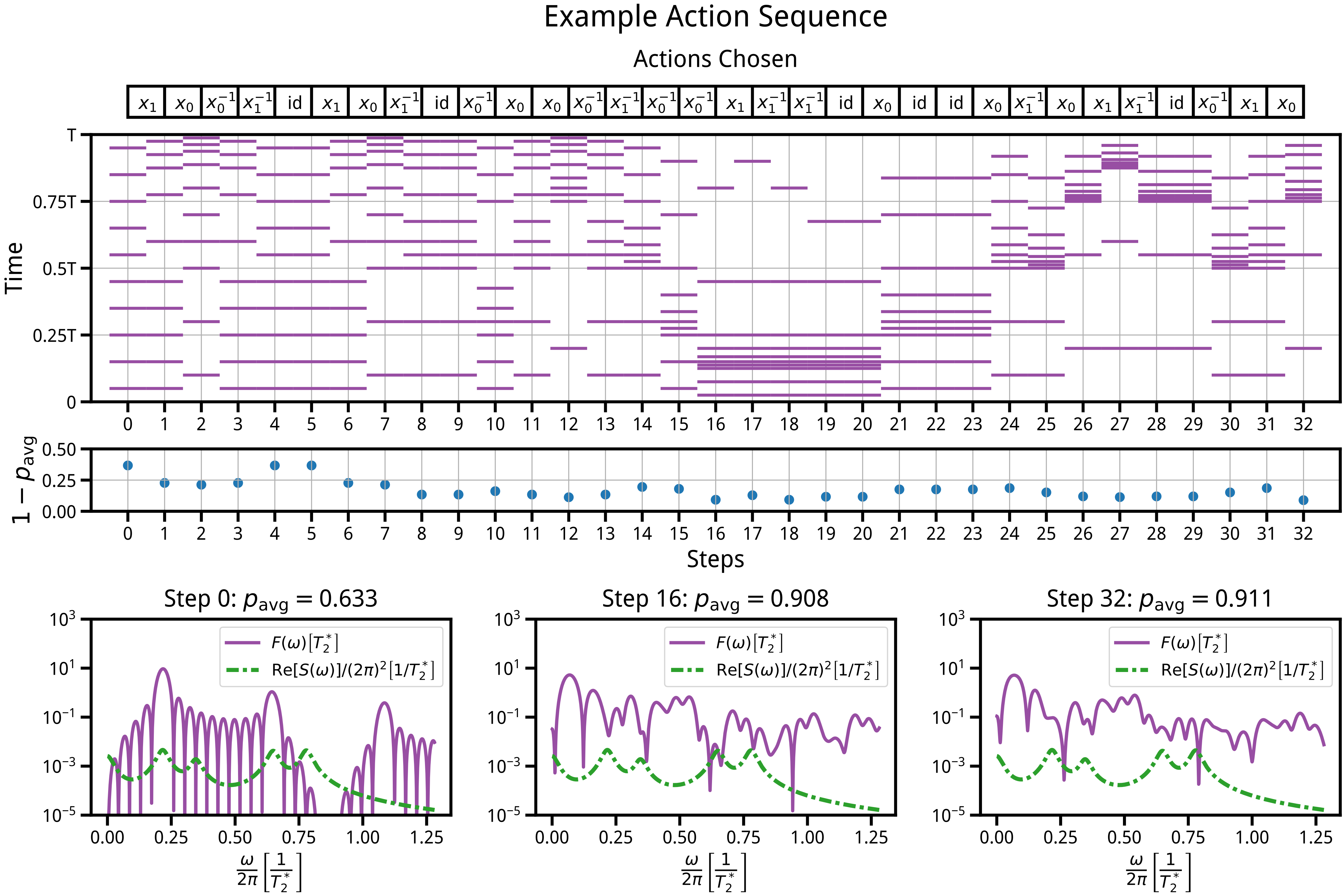}
    \caption{Example of an optimal action sequence found by the agent. From top to bottom, the panels show: (i) the action chosen by the agent at each step, (ii) the evolution of pulse times with each action, (iii) the corresponding values of $1-p_\mathrm{avg}$ at each step, and (iv) the filter functions at steps 0, 16, and 32 alongside the noise spectrum of this environment. The noise spectrum is a sum of randomly generated Lorentzian distributions, as described in Section \ref{sec:generation_spectra}. To express the filters and spectra in physically meaningful units of time and frequency, we scale by the free induction decay time $T_2^*$ [Eq.~(\ref{eq:T2star})]. In this example, the agent starts from a CPMG pulse sequence with $N=10$ pulses at step 0 and ultimately discovers a nontrivial, improved pulse sequence by step 32.}
    \label{fig:action-sequence}
\end{figure*}

As an example of the agent's ability to suppress one of these randomly generated noise spectra, we initialize it in a CPMG sequence with $N=10$ pulses and train it from scratch. 
Over the course of 5000 episodes, each requiring a single fidelity measurement, the agent discovers a pulse sequence that improves the fidelity [Eq.~(\ref{eq:avg_prob_of_zero})] from $p_\mathrm{avg} = 0.633$ to $p_\mathrm{avg} = 0.911$.
Fig.~\ref{fig:action-sequence} shows the sequence of actions the agent applied to generate this final, enhanced pulse sequence.

The top horizontal bar in Fig.~\ref{fig:action-sequence} displays the actions selected by the agent at each step. The two graphs below illustrate how the pulse times (shown as horizontal purple lines) and the corresponding infidelities, $1-p_\mathrm{avg}$ (blue dots), evolve throughout the episode. Although the plots show the infidelity at each step, the agent only computes $1-p_\mathrm{avg}$ and receives the associated reward after the final action (see Appendix \ref{app:our_agent_and_environment}). Despite this sparse feedback, the agent's capacity for long-term planning allows it to learn effectively. Because $1-p_\mathrm{avg}$ does not strictly decrease at every step, we can infer that this long-term strategy enables the agent to traverse local minima on its way to the final optimized solution.

The bottom three panels correspond to steps 0, 16, and 32. Each panel shows the filter function associated with the pulse sequence at that step alongside the noise spectrum the agent is attempting to suppress (without explicit knowledge of it). We can see that the agent has learned to minimize the overlap [Eq.~(\ref{eq:chi})] between the filter function and the noise spectrum by aligning some of the crests (troughs) of the filter function with the troughs (crests) of the noise spectrum. Thus, even without any direct knowledge of the underlying environment, the agent acquires implicit understanding through the rewards it receives.

\subsection{Consistency of Agent Performance}

\begin{figure}[!ht]
    \centering
    \includegraphics[width=\columnwidth]{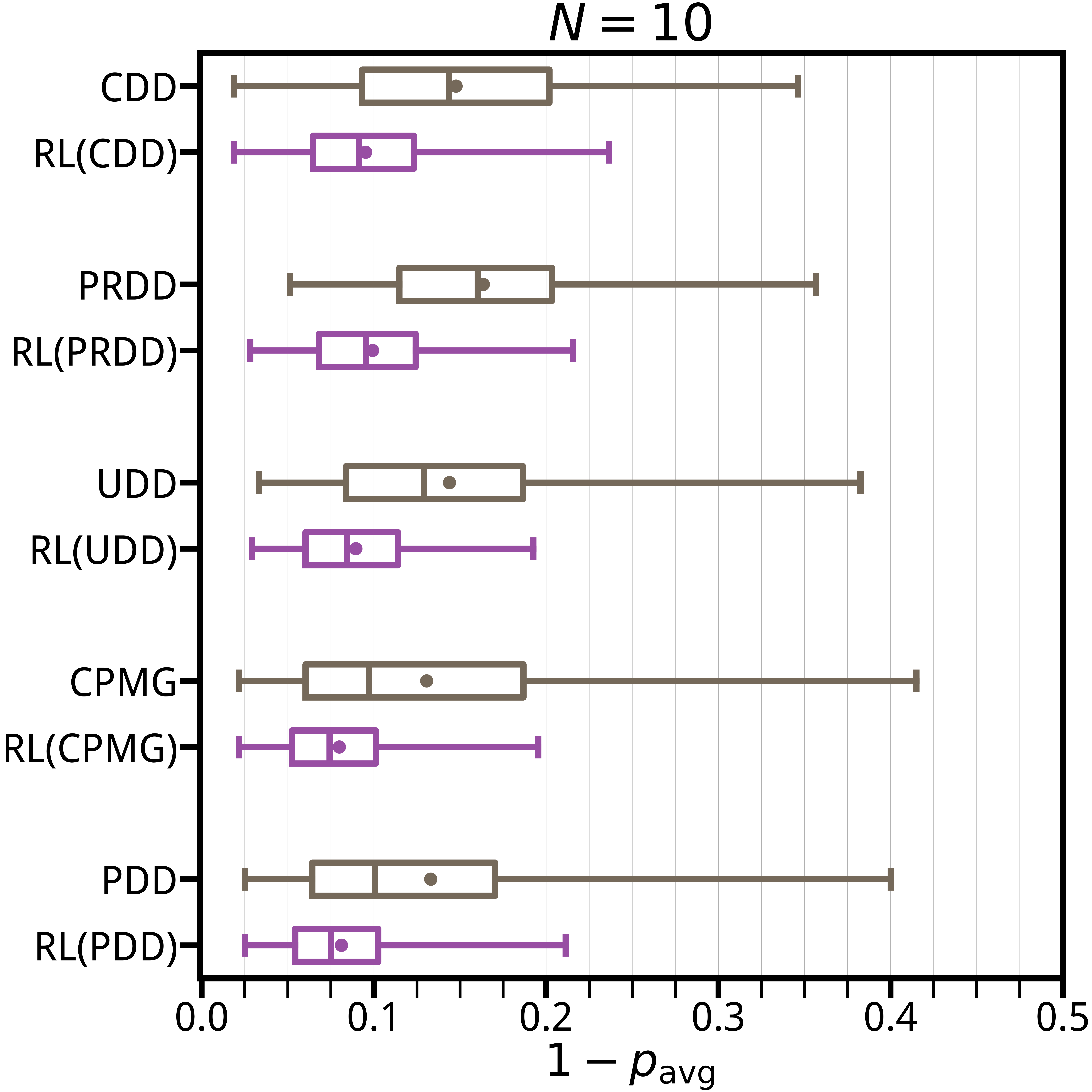}
    \caption{Comparison of initial and final pulse sequence performance using box-and-whisker plots. Initial sequences are shown in gray, while the best final sequences found by the agent after 5000 episodes are shown in purple, denoted as $\mathrm{RL}(\cdot)$, where $\cdot$ indicates the initial sequence. We test the agent on 200 randomly generated noise spectra $\{ S_i(\omega) \, | \, i=1,\ldots,200 \}$. For CDD, UDD, CPMG, and PDD, the agent is initialized in the same respective sequence for all episodes targeting a given $S_i(\omega)$ (CDD with $N=10$ corresponds to $\mathrm{CDD}_4$). For PRDD, a new random initial sequence is generated for each $S_i(\omega)$, but remains fixed across episodes for that spectrum. The box-and-whisker plot conventions are as follows: the whiskers indicate the minimum and maximum values of $1-p_\mathrm{avg}$, the vertical line inside the box shows the median, and the dot shows the mean. The box itself represents the interquartile range, with each section (left whisker, left box, right box, right whisker) corresponding to one quarter of the data.}
    \label{fig:fid-vs-seq-N10}
\end{figure}

To evaluate the consistency of the agent’s performance, we test it starting from five different initial pulse sequences: PDD, CPMG, UDD, CDD, and a new sequence we call Pseudorandom Dynamical Decoupling (PRDD). A PRDD sequence of $N$ pulses is constructed by dividing the evolution interval $[0, T]$ into $N$ equal subintervals and randomly selecting a pulse time within each subinterval. For each randomly generated noise spectrum, we calculate $p_\mathrm{avg}$ for both the agent’s initial and optimized pulse sequences. These results are summarized as box-and-whisker plots for each initial sequence and its corresponding agent-optimized sequence. For every spectrum and initial sequence, the agent was trained over 5000 learning episodes. 
The results are shown in Fig.~\ref{fig:fid-vs-seq-N10}. Starting from any initial pulse sequence, the RL agent reduces both the median and mean of $1-p_\mathrm{avg}$ by roughly a factor of 1.5, while decreasing the total range and interquartile range by approximately a factor of 2. Furthermore, the agent’s final performance is largely independent of the initial sequence, demonstrating consistent optimization across different starting conditions.

We note that we do not impose a constraint on the minimum pulse spacing, $\tau = \min(t_{j+1} - t_j)$, requiring it to exceed a global “minimum switching time” $\tau_\mathrm{min}$. While the relationship between $\tau$, $\tau_\mathrm{min}$, and the noise spectrum’s cutoff frequency $\omega_c$ is important for setting theoretical bounds on the best achievable performance for a given sequence \cite{Khodjasteh_LimitsPreservingQuantum_2011}, enforcing such a constraint is beyond the scope of this work. Our focus here is primarily a proof of concept for the proposed action set, allowing the agent to freely explore all potential advantages, including placing consecutive pulses arbitrarily close together. In future studies, we plan to extend the framework to more realistic pulse models, incorporating non-instantaneous, imperfect pulses with finite bandwidth and nonzero minimum switching times.

It is difficult to predict how the agent would perform for larger $N$. At first glance, one might hypothesize that for larger $N$, the agent would require a greater number of steps $M$ per episode to maintain strong performance; as $N$ increases, the spacing between pulses in the initial sequences decreases, necessitating the use of elements of Thompson's group~$F$ with more breakpoints and, consequently, more action function compositions (steps) within each episode to better approximate the optimal order-preserving homeomorphism $u$ between the initial sequence and the optimal sequence (see Sections \ref{sec:decision_problem} and \ref{sec:thompsons_group_f}). If this were the case, one would expect the performance of agents initialized in UDD (whose minimum interpulse spacing scales as $1/N^2$) to degrade more rapidly compared to agents starting in CDD, PRDD, CPMG, or PDD (whose minimum interpulse spacing scales as $1/N$). However, depending on the initial sequence and the structure of the noise spectrum, the function $u$ might not be complicated, requiring less breakpoints (and thus less steps) than expected. Of course, we do not know \textit{a priori} the noise spectrum nor the optimal map $u$, but using the smallest number of steps $M$ is important. Choosing an unnecessarily large $M$ would force the agent to make extra decisions in each episode before receiving a reward, potentially making it harder for the agent to learn. 

This question is further complicated by the fact that the agent's performance depends on the various hyperparameters of the agent's neural networks (see Appendix \ref{app:hyperparameters}). All of these considerations make predicting the agent's performance as a function of $N$ extremely difficult.
When using our method in practice, one can circumvent these issues by doing the following. First, choose $N$ according to experimental constraints.
Next, train an agent to reproduce the vast majority of pulse sequences in the search space of all valid pulse sequences of length $N$, using a reward based on the distance between the initial sequence and these target sequences in the search space. After tuning $M$ and and the neural hyperparameters to optimize this new agent's performance, use these tuned hyperparameters to do live learning for error suppression, doing further tuning as necessary.

\section{Conclusion and Outlook}

We have introduced a novel action set for reinforcement learning (RL) and sequential decision problems more broadly, consisting of five functions: the two generators of Thompson's group~$F$, their inverses, and the identity function. 
This action set applies to any sequential decision problem in which the state is represented by a bounded, ordered sequence of real numbers, $s=(t_1, \ldots, t_M)$ with $t_j \in [t_\mathrm{min}, t_\mathrm{max}]$ and $t_{j+1} > t_j$.
It is particularly suited for optimizing the timing of events in a time series. 
By distinguishing between the learning process's steps and the real-world arrow of time, the agent is able to do collective operations on the event times, eventually learning the optimal sequence of events.

We apply this action set to a classic quantum control problem, showing that an RL agent can consistently design dynamical decoupling sequences that suppress dephasing noise in a qubit more effectively than traditional sequences. 
The agent optimizes the $N$ pulse times of a sequence by adjusting all times across $[0,T]$ at each learning step. 
This converts what would otherwise be a costly $N$-dimensional, non-convex optimization problem into an $M$-step decision problem with only five choices per step, evaluating the objective function only after completing $M$ steps. 
While we maximize the ensemble-averaged probability $p_\mathrm{avg}$ of returning to the initial qubit state $\ket{0}$, this method could also optimize other quantum-state-dependent quantities, such as the classical or quantum Fisher information, making it relevant not only for error suppression, but also for quantum metrology.

Since our RL algorithm is model-free, we expect it to be robust against pulse errors in experimental implementations, and future work will extend this study to non-ideal pulses. Here, assuming classical noise allows us to frame the problem as a deterministic Markov decision process. 
Introducing classically random pulse errors or statistical estimates of $p_\mathrm{avg}$ would generalize it to a stochastic Markov decision process, where DDQN should remain effective given existing convergence guarantees for Q-learning \cite{Watkins_LearningDelayedRewards_1989,Watkins_Qlearning_1992}. 
Including mixed-state dynamics or non-Gaussian classical noise similarly leads to a stochastic process.

However, introducing non-classical (quantum) noise, where noise terms do not commute at all times, could introduce information backflow between the system and bath, making the process non-Markovian. 
Specifically, if the system’s response after measurement and reinitialization depends on prior sequences, this would indicate memory effects carried by the bath \cite{Pollock_OperationalMarkovCondition_2018,Pollock_NonMarkovianQuantumProcesses_2018}. 
In such cases, one could augment the RL state to include pulse sequences from the previous $k$ episodes, at the cost of increasing state-space size and neural network complexity. This overhead could be mitigated by imposing temporal symmetry ($t_{N-j}=t_j$) on sequences, and in practical systems, this restriction on the search space likely retains enough optima for the agent to exploit.
Alternatively, deep Q-learning with recurrent architectures, such as Long Short-Term Memory (LSTM) units, could address non-Markovian dynamics by retaining long-term trends while responding to short-term fluctuations \cite{Hausknecht_DeepRecurrentQLearning_2015,Hochreiter_LongShortTermMemory_1997}.

Many quantum systems also exhibit noise drift, where the spectrum $S(\omega)$ changes over time \cite{Klimov_FluctuationsEnergyRelaxationTimes_2018}. If the agent can learn an optimal sequence faster than the drift timescale, our method remains viable. Retraining can be performed periodically with minor adjustments to the agent’s neural network weights, initializing from the previous optimal sequence and network, thus reducing the number of episodes required to adapt to new noise conditions.
Future work includes closed-loop learning off of an experimental implementation with nitrogen-vacancy center qubits. These qubits promise to be a good first demonstration of our live learning method since their coherence times are often limited by dephasing ($T_2<T_1$) \cite{Segawa_NanoscaleQuantumSensing_2023}, and DD has already been shown to improve their dephasing times at room temperature \cite{Bar-Gill_SuppressionSpinbathDynamics_2012}.

\begin{acknowledgments}
The authors thank William Schenken, John Wilson, Jarrod Reilly, and Shah Saad Alam for helpful discussions. We acknowledge support from NSF OMA 2016244 and NSF PFC 2317149. RL methods were developed with support from NASA under grant number 80NSSC23K1343. CM and MH
also acknowledge support from the Department of Energy Quantum Systems Accelerator, Grant No.\ 7565477. SS~acknowledges support from the NSF CAREER Award 2443684 and a Sloan Research Fellowship.
\end{acknowledgments}

\appendix

\section{\label{app:filter_function_formalism}Filter Function Formalism}

There are several established approaches for quantifying qubit decoherence using filter-function techniques (see, e.g., \cite{Cywinski_HowEnhanceDephasing_2008, Uhrig_ExactResultsDynamical_2008, Biercuk_DynamicalDecouplingSequence_2011, Paz-Silva_GeneralTransferFunctionApproach_2014, Chalermpusitarak_FrameBasedFilterFunctionFormalism_2021}). Here, we describe the specific physical model used in our simulations, with the aim of rederiving Eqs.~(\ref{eq:avg_prob_of_zero}) and (\ref{eq:chi}). We consider a general qubit state
\begin{equation}
\ket{\theta, \phi} = \cos(\theta / 2) \ket{0} + e^{i\phi} \sin(\theta / 2)\ket{1}\,,
\end{equation}
where $\theta$ and $\phi$ denote the polar and azimuthal angles on the Bloch sphere, respectively. The qubit is subjected to a pure-dephasing Hamiltonian
\begin{equation}
\hat{H}(t) = \frac{\hbar}{2} \left[ \Omega + \beta(t) \right] \hat{\sigma}_z \, ,
\end{equation}
where $\Omega$ is the qubit's natural frequency, $\beta(t)$ is a zero-mean, Gaussian, wide-sense stationary stochastic process representing frequency noise, and $\hat{\sigma}_z$ is the Pauli-$Z$ operator. 

The effect of dephasing is greatest when the qubit is on the equator of the Bloch sphere, \textit{i.e.}, when $\theta = \pi/2$. To study the worst-case performance of our noise suppression scheme, we initialize the qubit in the $\ket{0}$ state and apply a $\hat{R}_y(\pi/2)$ rotation at $t=0$, placing the state on the equator, specifically in the $\ket{+}$ state. After interspersing free evolution under $\hat{H}(t)$ with ideal $\hat{R}_x(\pi)$ pulses, we apply a $\hat{R}_y(-\pi/2)$ rotation at final time $T$ and measure the probability $p$ of measuring $\ket{0}$.

Using the Born rule, we find
\begin{equation}
    p = \left|\left<0 \left| \hat{R}_y\left(-\frac{\pi}2 \right) \right| \frac{\pi}2, \phi(T)\right>\right|^2 = \frac{1 + \cos\bigl(\phi(T)\bigr)}2\,.    
\end{equation}
 However, the phase accumulated $\phi(T)$ depends on the classical noise $\beta(t)$ and thus $\phi(T)$ is itself a random variable. Consequently, we must calculate the average value of $p$ by averaging over the noise realizations, so
\begin{equation}
    p_\mathrm{avg} = \frac{1 + \langle \cos\bigl(\phi(T)\bigr)\rangle}2 .
    \label{eq:p_cosine}
\end{equation}
We express the final phase $\phi(T)$ of the qubit in terms of the initial time $t_0 = 0$, the pulse times $(t_1, \ldots, t_N)$, and the final time $t_{N+1} = T$. To do so, we move to the rotating frame of the qubit's natural frequency. For $n = 1, \ldots, N+1$, let $\hat{U}_n$ be the free evolution operator that propagates the state from time $t_{n-1}$ to $t_n$ under the rotating frame Hamiltonian
\begin{equation}
    \hat{H}_{RF} = \frac{\beta(t)}2 \hat{\sigma}_z\,.
\end{equation}
The total evolution operator from $t=t_0=0$ to $t=t_{N+1}=T$ is
\begin{equation}
    \hat{U}_{N+1} \hat{R}_x(\pi) \hat{U}_N \ldots \hat{U}_2 \hat{R}_x(\pi) \hat{U}_1\,.
\end{equation}
Noting that each $\hat{U}_n$ contributes a phase $\int_{t_{n-1}}^{t_n} \beta(t')\, dt'$ to the qubit and that $\hat{R}_x(\pi)$ flips the sign of the accumulated phase, one can show by induction, for a qubit initialized with $\phi=0$ at $t=0$,
\begin{eqnarray}
    \phi(T) &=& (-1)^N \Biggl[ \, \sum_{k=0}^N (-1)^k \int_{t_k}^{t_{k+1}} \beta(t') \, dt' \Biggr] \nonumber \\
    &=& (-1)^N \Biggl[\int_0^T \beta(t') \times  \\
    &&\quad \left( \sum_{k=0}^N (-1)^k \left( \Theta(t' - t_k) - \Theta(t' - t_{k+1}) \right) \right) dt' \Biggr], \nonumber
\end{eqnarray}
where $\Theta(t)$ is the Heaviside step function. If we define the switching function 
\begin{equation}
    f(t, t_1, \ldots, t_N, T) \vcentcolon= \sum_{k=0}^N (-1)^k \left( \Theta(t - t_k) - \Theta(t - t_{k+1}) \right) ,
    \label{eq:switching_function}
\end{equation}
then
\begin{equation}
    \phi(T) = (-1)^N \left[\int_{-\infty}^\infty \beta(t') f(t', t_1, \ldots, t_N, T) \, dt'\right] ,
    \label{eq:phase_accumulated}
\end{equation}
where we have used that $f(t', t_1, \ldots,t_N, T) = 0$ for $t' < 0$ and $t' > T$.

Using Eq.~(\ref{eq:phase_accumulated}), we note that because $\beta(t')$ is a random Gaussian variable, $\phi(T)$ is also Gaussian. Moreover, since $\langle \beta(t') \rangle = 0$ for all $t'$, it follows that $\langle \phi(T) \rangle = 0$. We can now simplify Eq.~(\ref{eq:p_cosine}). Expanding the cosine via its Taylor series gives
\begin{equation}
    \langle \cos(\phi(T)) \rangle = \sum_{j=0}^\infty \frac{(-1)^j}{(2j)!} \langle \phi(T)^{2j} \rangle\,.
\end{equation}
Using the Gaussian moment theorem (see Section 1.6.1 of \cite{Mandel_OpticalCoherenceQuantum_1995}), we have
\begin{eqnarray}
     \langle \phi(T)^m \rangle &&= 
     \langle \left( \phi(T) - \langle \phi(T) \rangle \right)^m \rangle \nonumber \\
     &&=\begin{cases}
         (m - 1)!! \langle \phi(T)^2 \rangle^{m/2}, & m \text{ even} \\
         0, & m \text{ odd ,}
     \end{cases}
     \label{eq:moments_of_phi}
\end{eqnarray}
where $(m-1)!!$ denotes the double factorial of $m-1$.
Substituting Eq.~(\ref{eq:moments_of_phi}) into the Taylor series for $\langle \cos(\phi(T)) \rangle$, noting that $(2j-1)!!/(2j)! = 1/(2j)!!=1/(2^j(j!))$, and recognizing the resulting exponential series, we obtain
\begin{equation}
    p_\mathrm{avg}=\frac{1}{2}\left(1 + e^{-\langle \phi(T)^2\rangle/2}\right) .
    \label{eq:p_exponential}
\end{equation}
Calculating $\langle \phi(T) \rangle$ directly from Eq.~(\ref{eq:phase_accumulated}) in the time domain does not make transparent how the pulse times $(t_1, \ldots, t_N)$ suppress the effect of the noise $\beta(t)$. The structure becomes much clearer upon transforming to Fourier space. Denote the Fourier transform of a function $g(t)$ by
\begin{equation}
    \mathcal{F}[g(t)](\omega) = \int_{-\infty}^{\infty} g(t) e^{i\omega t} \, dt\,.
\end{equation}
Using Eq.~(\ref{eq:somega}) and the fact $\beta(t)$ is wide-sense stationary, we have
\begin{eqnarray}
    \langle \beta(t') \beta(t'') \rangle &=& \langle \beta(t' - t'') \beta(0) \rangle \nonumber\\ &=& \mathcal{F}^{-1}[S(\omega)](t'-t'') \nonumber \\ &=& \frac{1}{2\pi} \int_{-\infty}^{\infty} S(\omega) e^{-i\omega(t'-t'')} \, d\omega \,.
\end{eqnarray}
Because $\beta$ and $f$ are real-valued, the identities $S(-\omega) = S(\omega)^*$ and \begin{eqnarray}
    &&\mathcal{F}[f(t, t_1, \ldots, t_N, T)](-\omega, t_1, \ldots,t_N,T) = \nonumber\\
    && \quad \mathcal{F}[f(t,t_1, \ldots, t_N,T)](\omega, t_1, \ldots, t_N, T)^*
\end{eqnarray}
hold. Defining $\chi\vcentcolon=\langle \phi(T)^2 \rangle/2$ as the attenuation function, we obtain
\begin{equation}
    \chi = \frac{1}{2\pi} \int_{0}^{\infty} \mathrm{Re}[S(\omega)] F(\omega, t_1, \ldots, t_N, T) \, d\omega \,,
    \label{eq:chi_rederived}
\end{equation}
where the filter function $F$ is 
\begin{eqnarray}
&&F(\omega, t_1, \ldots, t_N, T) = \nonumber\\
&&\quad \left|\mathcal{F}[f(t, t_1, \ldots, t_N, T)](\omega, t_1, \ldots,t_N,T)]\right|^2 .
\end{eqnarray}

This frequency domain expression is much easier to interpret than the time domain form of $\langle \phi(T)^2 \rangle / 2$. Taken together, Eqs.~(\ref{eq:p_exponential}) and (\ref{eq:chi_rederived}) show that qubit coherence is maximized when the overlap between $\mathrm{Re}[S(\omega)]$ and $F(\omega, t_1, \ldots, t_N, T)$ is minimized. More formally, $\chi$ is the $L_2$ inner product between the functions $\mathrm{Re}[S(\omega)]$ and $F(\omega, t_1,\ldots,t_N,T)$ over the positive frequency domain. Here, coherence is maximized when this inner product is minimized.

We also introduce a useful timescale, the free-induction decay time $T_2^*$, which characterizes the overall magnitude of the noise spectrum $S(\omega)$. In a free-induction decay (FID) experiment, where no pulses are applied and the total evolution time is $T_\mathrm{FID}$, the corresponding filter function is
\begin{equation}
    F_\mathrm{FID}(\omega) = T_\mathrm{FID}^2\mathrm{sinc}^2\left(\frac{\omega T_\mathrm{FID}}2\right) ,
\end{equation}
where $\mathrm{sinc}(x):=\sin(x)/x$.

For most physically realistic noise spectra, there is an ultraviolet cutoff frequency $\omega_c$ such that $|\mathrm{Re}[S(\omega)]| \approx 0$ for $\omega>\omega_c$. When $T_\mathrm{FID} \ll 2 / \omega_c$, the filter function is approximately constant
\begin{equation}
    F_\mathrm{FID}(\omega) \approx T_\mathrm{FID}^2\,.
\end{equation}
Under this approximation, the attenuation becomes $\chi \approx (T_\mathrm{FID}/T_2^*)^2$, where
\begin{equation}
    T_2^* = \left( \frac{1}{2\pi} \int_0^{\omega_c} \mathrm{Re}[S(\omega)] \, d\omega \right)^{-1/2} .
    \label{eq:T2star}
\end{equation}

\section{\label{app:MathOfQLearning}The Mathematics of Q-Learning}

\subsection{Markov Decision Processes}
Before introducing Q-learning, we briefly review Markov decision processes (MDPs), the class of problems for which Q-learning was originally developed~\cite{Watkins_LearningDelayedRewards_1989, Watkins_Qlearning_1992}. An MDP consists of a set of states $\mathcal{S}$, a set of available actions $\mathcal{A}(s)$ for each state $s\in\mathcal{S}$, a set of numerical rewards $\mathcal{R} \subset \mathbb{R}$, and a (potentially probabilistic) policy function $\pi$ that specifies an action to take in each state~\cite{Sutton_ReinforcementLearningIntroduction_2018, Puterman_MarkovDecisionProcesses_1994}. Each action $a\in \mathcal{A}(s)$ induces a (possibly probabilistic) transition from one state to another, and the reward $r(s,a,s') \in \mathcal{R}$ received for taking action $a$ in state $s$ and arriving in state $s'$ may itself be a discrete or continuous random variable.

Using the language of reinforcement learning  (RL), we interpret $\mathcal{S}$, $\mathcal{A}$, and $\mathcal{R}$ as defining a possibly stochastic \textit{environment} with which a decision making  \textit{agent} interacts. The agent's behavior is governed by its \textit{policy} $\pi$.

For simplicity, we restrict our discussion to discrete, finite MDPs, in which $\mathcal{S}$ and each $\mathcal{A}(s)$ are finite sets, and state transitions occur in discrete time steps. It is often helpful to visualize an MDP as a directed, weighted, and colored graph: the states $\mathcal{S}$ form the nodes, the actions $\mathcal{A}(s)$ for each state correspond to directed edges, the state-transition probabilities serve as edge weights, and the rewards label the edges as colors. The fact that the available action set $\mathcal{A}(s)$ may differ from state to state simply reflects that each node in the graph may have a different number of outgoing edges.

At step $k$, the MDP occupies a specific state $s_k$, selects an action $a_k \in \mathcal{A}(s_k)$ according to its policy, and with probability $p(s_{k+1}, r_{k+1} | s_k, a_k)$, the agent receives a reward $r_{k+1}$ and transitions to a new state $s_{k+1}$. The fact that this probability depends only on the current state $s_k$ and action $a_k$  (and not on earlier history) is known as the \textit{Markov property}. Decision processes that do not satisfy this property are called \textit{non-Markovian}. The probability of choosing action $a_k$ when in state $s_k$ under policy $\pi$ is denoted $\pi(a_k|s_k)$.

Although we may deterministically initialize the MDP in a specific state $s_0 \in \mathcal{S}$, we generally cannot predict with certainty the subsequent state $s_1$ or reward $r_1$. This uncertainty arises from the stochasticity inherent both in action selection and in state transitions.
Consequently, we treat states, actions, and rewards as random variables.

We denote the random variables representing the state and action at step $k$ by $S_k$ and $A_k$, respectively, while the actual state and action realized in a given sample are written as $s_k$ and $a_k$. Similarly, the random variable corresponding to the reward received at step $k+1$, resulting from being in state  $S_k$ and taking $A_k$, is $R_{k+1}$, and the actual reward received by the agent is $r_{k+1}$.

Viewed as a graph, an $M$-step, nondeterministic \textit{trajectory} through the graph is itself a random variable, which can be represented as the sequence
\begin{equation}
    (S_0, A_0, R_1,S_1,\ldots,S_{M-1},A_{M-1},R_M,S_M)\,.
\end{equation}
In general, we can consider infinitely long trajectories ($M \rightarrow \infty$), corresponding to control tasks that may never terminate. Assuming an infinite horizon is mathematically convenient, and tasks that terminate after a finite number of steps can be reformulated as infinite-horizon trajectories by introducing \textit{terminal states}.

A terminal state (also called an ``absorbing state'') is a special state that deterministically transitions only to itself and receives zero reward for doing so. Using terminal states, all finite-horizon tasks (with a known maximum number of steps) and indefinite-horizon tasks (with an unknown maximum number of steps) can be expressed as infinite-horizon tasks by augmenting the MDP with additional states that track the number of steps~\cite{Sutton_ReinforcementLearningIntroduction_2018}. Because of this equivalence, we adopt the infinite-horizon notation in this summary, taking $M \rightarrow \infty$. Moreover, Q-learning algorithms typically learn by completing a series of episodes, each of which are a finite-horizon or indefinite-horizon trajectory. However, the results here are true for each episode of Q-learning, so we make no distinction between episodes in our notation.

The objective in an MDP is to maximize the expected \textit{return}~\cite{Sutton_ReinforcementLearningIntroduction_2018}. 
The return $G_i$ is the ``cumulative discounted reward'' available after step $i$ of the random trajectory.
It is a random variable defined as
\begin{eqnarray}
    G_i &&\vcentcolon= R_{i+1} + \gamma R_{i+2} + \gamma^2 R_{i+3} + \ldots , \nonumber \\
    &&=\sum_{k=i+1}^\infty \gamma^{k-i-1} R_k\,,
    \label{eq:return}
\end{eqnarray}
where $\gamma \in [0,1]$ is the \textit{discount factor}.

The discount factor $\gamma$ serves two key purposes. First, it tunes the agent's balance between long-term planning and short-term gratification: values of $\gamma$ close to 1 encourage the agent to prioritize long-term rewards, while values closer to 0 emphasize immediate rewards. Second, in infinite-horizon tasks, setting $\gamma<1$ ensures that the return $G_i$ converges mathematically, making the expected return finite and well-defined.

The policy function determines the agent’s choice of action at each step. A deterministic policy is a mapping $\pi: s \mapsto a$  that assigns a specific action to each state. More generally, a stochastic policy maps states to probability distributions over actions, so that the probability of taking action $a$ in state $s$ is $\pi(a|s)$. Using an initial state $s_0$ and a policy $\pi$, one can iteratively generate a trajectory through the state space.

The performance of a policy $\pi$ is quantified by the \textit{state-value function} $v_\pi(s)$, which gives the expected return when starting in state $s$ and subsequently following $\pi$. Formally, the value of state $s$ at step $i$ is
\begin{equation}
    v_\pi(s) \vcentcolon = \mathbb{E}_{\pi} \left[ G_i | S_i=s \right],
    \label{eq:state_value_function}
\end{equation}
where the expectation is taken over all possible trajectories generated by $\pi$ starting from $s$, either until reaching a terminal state or infinitely far into the future. The specific step $i$ at which $v_\pi(s)$ is evaluated is irrelevant; what matters is the expected return available from $s$.

The goal in an MDP, and in reinforcement learning more generally, is to find the optimal policy $\pi_*$ that maximizes the value $v_{\pi_*}(s)$  for every possible initial state $s_0$. 

\subsection{\label{app:Bellman}Bellman's Principle of Optimality}
The number of trajectories contributing to the expectation in the state-value function, Eq.~(\ref{eq:state_value_function}), is typically extremely large, since each action in a Markov decision process can lead to multiple possible next states. As a result, brute-force optimization of an entire sequence of $M$ actions (with $M$ potentially arbitrarily large) for every possible initial state is generally intractable. Reinforcement learning overcomes this difficulty by leveraging a key idea from dynamic programming: \textit{Bellman's principle of optimality}~\cite{Bellman_DynamicProgramming_2003}.

This principle asserts that, regardless of the initial state $s_0$ and the first action $(a_0)_*$ chosen by the optimal policy $\pi_*$, the remaining actions $(a_1)_*, \ldots, (a_{M-1})_*$ constitute an optimal policy relative to the next state~$(S_1)_*$. Intuitively, in the deterministic case, the optimal $(M-1)$-step trajectory starting from $(s_1)_*$ is embedded within the original $M$-step optimal trajectory starting from $s_0$. More generally, this recursive property holds not only for $(S_1)_*$ but for every state $(S_k)_*$ along the $M$-step optimal trajectory.

To formalize this idea, let us assume for simplicity that the reward $r(s,a,s')$ received for being in state $s$, taking action $a$, and transitioning to state $s'$ is deterministic. By the definition of the return, we have the recursive relation
\begin{equation}
    G_i = R_{i+1} + \gamma G_{i+1} ,
    \label{eq:return_recursion}
\end{equation}
which can be used to derive a recursion for the state-value function:
\begin{widetext}
\begin{eqnarray}
    v_\pi(s) &&= \mathbb{E}_\pi[R_{i+1} + \gamma G_{i+1} | S_i=s] , \nonumber \\
    &&=\sum_{a\in\mathcal{A}(s)} \pi(a|s) \sum_{s'\in\mathcal{S}} p(s'|s,a) \bigl[ r(s,a,s')+ \gamma \mathbb{E}_\pi[G_{i+1}|S_{i+1}=s'] \bigr] ,\nonumber \\
    &&= \sum_{a\in\mathcal{A}(s)} \pi(a|s) \sum_{s'\in\mathcal{S}} p(s'|s,a) \bigl[ r(s,a,s')
    + \gamma v_\pi(s') \bigr].
    \label{eq:state_value_recursion}
\end{eqnarray}
\end{widetext}
Equation~(\ref{eq:state_value_recursion}) is known as the {\em Bellman equation} for the state-value function, expressing the value of a state in terms of the immediate reward plus the discounted value of successor states.

Combining Bellman's principle of optimality with the recursion relation in Eq.~(\ref{eq:state_value_recursion}), we see that it is unnecessary to solve for the entire optimal action sequence at once. Instead, if the transition probabilities $p(s'|s,a)$, the rewards $r(s,a,s')$, and the optimal value function $v_*$ were known for a sufficiently large subset of states, we could iteratively reconstruct $v_*$ for all states in $\mathcal{S}$ using Eq.~(\ref{eq:state_value_recursion}). This process of refining a function estimate using current estimates of the same function is known as \textit{bootstrapping}.

Of course, in general we do not know $v_*$ for any state \textit{a~priori}. Classical dynamic programming (DP) addresses this by initializing value estimates arbitrarily and repeatedly applying the recursion in Eq.~(\ref{eq:state_value_recursion}), eventually converging to the optimal values~\cite{Sutton_ReinforcementLearningIntroduction_2018}. However, DP assumes full knowledge of the transition probabilities and rewards~\cite{Sutton_ReinforcementLearningIntroduction_2018}, which is typically unavailable in reinforcement learning. As a result, DP is suited for planning an optimal policy rather than learning one from interaction with an unknown environment.

\subsection{Q-Learning \label{app:Qlearning}}
To learn an optimal policy without knowledge of the environment’s dynamics, it is not sufficient to know which states yield the highest expected return; we also need to know how to reach those states. By evaluating how taking a specific action in a given state affects the expected return, we can train an agent to navigate state space optimally through a sequence of appropriate actions.

Given a policy $\pi$, the \textit{action-value function} $q_\pi(s,a)$ is defined as the expected return when starting in state $s$, taking action $a$, and then following $\pi$ thereafter~\cite{Sutton_ReinforcementLearningIntroduction_2018}: \begin{equation}
    q_\pi(s,a) \vcentcolon = \mathbb{E}_{\pi} \left[ G_i | S_i=s, A_i=a \right]\,.
    \label{eq:action_value_function}
\end{equation}
Using the recursive definition of the return Eq.~(\ref{eq:return_recursion}), we can derive a recursion for action values in terms of state values:
\begin{eqnarray}
    q_\pi(s,a) &&= \mathbb{E}_\pi[R_{i+1} + \gamma G_{i+1} | S_i=s, A_i=a] , \nonumber \\
    &&= \sum_{s'\in\mathcal{S}} p(s'|s,a) \bigl[ r(s,a,s')+ \gamma \mathbb{E}_\pi[G_{i+1}|S_{i+1}=s'] \bigr] ,\nonumber \\
    &&= \sum_{s'\in\mathcal{S}} p(s'|s,a) \bigl[ r(s,a,s')
    + \gamma v_\pi(s') \bigr]\,.
    \label{eq:action_value_recursion}
\end{eqnarray}
Comparing this with Eq.~(\ref{eq:state_value_recursion}), the state-value function can be expressed in terms of action values as
\begin{equation}
    v_\pi(s) = \sum_{a \in \mathcal{A}(s)} \pi(a|s) q_\pi(s,a).
    \label{eq:state_value_terms_of_action_value}
\end{equation}
This shows that $v_{\pi}(s)$ is a linear function over the probability simplex defined by $\pi(a|s)$. Its maximum occurs at one of the vertices of the simplex, i.e., when $\pi$ deterministically selects the action with the largest action value. Such a policy is referred to as being \textit{greedy}. Using the definition of the optimal state-value function $v_*(s) \vcentcolon= \max_\pi v_\pi(s)$ , we obtain
\begin{equation}
    v_*(s) = \max_{a \in \mathcal{A}(s)} q_*(s, a) , 
    \label{eq:optimal_state_value_terms_optimal_action_value}
\end{equation}
where $q_*(s,a) \vcentcolon= \max_\pi q_\pi(s,a)$. This implies that the optimal state-value function is fully determined by the optimal action-value function, and the optimal policy is deterministic:
\begin{equation}
    \pi_*(a|s) = 
    \begin{cases}
        1, &\text{ if } a=\argmax_{a'\in\mathcal{A}(s)} q_*(s,a') \\
        0, &\text{ otherwise. }
    \end{cases}
    \label{eq:optimal_policy_cases}
\end{equation}
Equation~(\ref{eq:optimal_state_value_terms_optimal_action_value}) shows that learning the optimal state values is equivalent to learning the optimal action values. Combining Eqs.~(\ref{eq:action_value_recursion}) and (\ref{eq:optimal_state_value_terms_optimal_action_value}) gives the Bellman optimality equation for action values:
\begin{equation}
    q_*(s,a) = \sum_{s'\in\mathcal{S}} p(s'|s,a) \Bigl[ r(s,a,s')
    + \gamma \max_{a' \in \mathcal{A}(s')} q_*(s', a') \Bigr].
    \label{eq:Bellman_opt_eq_action_vals}
\end{equation}
The remaining question is: how can one learn the optimal action-value function $q_*$ when the transition probabilities and rewards are unknown?

At first glance, one might consider using Monte Carlo methods, which estimate expected returns by randomly generating action sequences and iteratively updating their estimates of the optimal action values. However, Monte Carlo methods must wait until the end of each episode to update their policy, and therefore cannot exploit the benefits of bootstrapping. In contrast, temporal-difference (TD) methods, such as Q-learning, combine the strengths of both approaches: they learn directly from experience like Monte Carlo methods while also using bootstrapping to update estimates incrementally. As a result, TD methods are typically more time- and space-efficient and often converge to the optimal value function faster in practice~\cite{Sutton_ReinforcementLearningIntroduction_2018}.

Q-learning algorithms begin by arbitrarily initializing \textit{estimates} of $q_*(s,a)$, called \textit{Q-values} and denoted by $Q(s,a)$. The goal of Q-learning is to iteratively update these estimates~$Q(s,a)$ so that they converge to the optimal action-value function~$q_*(s,a)$ for every state $s$ and action $a$. Unlike Monte Carlo methods, Q-learning updates its estimates $Q$ after each step rather than waiting until the end of an episode. To balance exploration and exploitation, Q-learning can use two (potentially distinct) policies: a target policy $\pi$ and a behavior policy $b$. It is an \textit{off-policy} algorithm, meaning that it uses experience generated under the behavior policy $b$ to improve its target policy $\pi$ toward the optimal policy $\pi_*$.

We define a greedy, deterministic target policy $\pi$ in the same manner as Eq.~(\ref{eq:optimal_policy_cases}), but using our current estimates $Q(s,a)$:
\begin{equation}
    \pi(a|s) = 
    \begin{cases}
        1, &\text{ if } a=\argmax_{a'\in\mathcal{A}(s)} Q(s,a') \\
        0, &\text{ otherwise. }
    \end{cases}
    \label{eq:Qlearning_target_policy}
\end{equation}
To encourage exploration, the behavior policy $b$ is chosen to be $\varepsilon$-greedy. At each step, with probability $1-\varepsilon$, the agent selects the action with the highest estimated Q-value; with probability $\varepsilon$, it selects an action uniformly at random from $\mathcal{A}(s)$. Formally, the behavior policy can be written as:
\begin{equation}
    b(a|s) = 
    \begin{cases}
        1 - \varepsilon + \frac{\varepsilon}{|\mathcal{A}(s)|}, &\text{ if } a=\argmax_{a'\in\mathcal{A}(s)} Q(s,a') \\
        \frac{\varepsilon}{|\mathcal{A}(s)|}, &\text{ otherwise.}
    \end{cases}
    \label{eq:Qlearning_behavior_policy}
\end{equation}
The choice of $\varepsilon$ depends on the specific reinforcement learning problem. While keeping $\varepsilon$ constant is simple, it is often advantageous to start with $\varepsilon=1$  and decay it over episodes or based on agent performance. This strategy promotes exploration early in training while favoring exploitation later, when the Q-values are more reliable.

A Q-learning agent ``learns'' by updating its Q-values based on the environment’s response to its actions. At the $i$-th step, if the agent is in state $s_i$, takes action $a_i$, receives reward $r_{i+1}$, and observes the next state $s_{i+1}$, its Q-values are updated according to:
\begin{equation}
    Q(s_i,a_i) \leftarrow Q(s_i, a_i) + \alpha (Y(r_{i+1}, s_{i+1}) - Q(s_i,a_i))\,,
    \label{eq:Qlearning_update_rule}
\end{equation}
where $\alpha \in (0,1)$ is the learning rate, controlling how rapidly the Q-values move toward the target, and
\begin{equation}
    Y(r_{i+1}, s_{i+1}) \vcentcolon= r_{i+1} + \gamma \max_{a' \in \mathcal{A}(s_{i+1})} Q(s_{i+1}, a')
    \label{eq:Qlearning_target}
\end{equation}
is the \textit{target} at step $i$. Larger values of $\alpha$ cause $Q(s_i, a_i)$  to converge more quickly toward the target $Y(s_i, a_i)$.

Rewriting the Bellman optimality equation for action values Eq.~(\ref{eq:Bellman_opt_eq_action_vals}) in a more general form:
\begin{equation}
    q_*(s,a)=\mathbb{E}_{\pi_*}\left[ R_{i+1} + \gamma \max_{a' \in \mathcal{A}(s')} q_*(s',a') \, \Bigg| \, S_i=s, A_i=a\right] ,
    \label{eq:generalized_Bellman_opt_eq_action_vals}
\end{equation}
we see that Eq.~(\ref{eq:Qlearning_target}) closely resembles Eq.~(\ref{eq:generalized_Bellman_opt_eq_action_vals}) but with two key differences: it uses the agent's current Q-value instead of the true optimal values $q_*$, and it is conditioned on the observed reward $R_{i+1}=r_{i+1}$ and next state $S_{i+1}=s_{i+1}$. Like $q_*$, the target $Y$ is inherently greedy, selecting the follow-up action that the agent currently estimates to be best. The difference between the target and the current action-value estimate $Y(r_{i+1}, s_{i+1}) - Q(s_i,a_i)$ is known as the \textit{temporal difference error}.

It was proven~\cite{Watkins_LearningDelayedRewards_1989, Watkins_Qlearning_1992} that for stochastic, stationary MDPs, where the state transition probabilities do not change over time, with bounded rewards and a properly decaying learning rate $\alpha$, the Q-values $Q$ converge to the true optimal action values $q_*$ in the limit where each state-action pair is visited infinitely often. This result assumes that both the state and action spaces are discrete and finite, and that $Q$ is represented as a lookup table, with an entry $Q(s,a)$ for every state $s$ and action~$a$.

In practice, however, the state space can be so large that a table-based representation becomes infeasible, as was the case in our work.

\subsection{Deep Q-Learning}
A Deep Q-Network (DQN) addresses this challenge by approximating the action-value function $Q(s,a)$ using a neural network~\cite{Mnih_HumanlevelControlDeep_2015}. In supervised learning, neural networks are trained on labeled data: the network's parameters (its weights and biases) are adjusted so that, after sufficient training, the network approximates the true function mapping inputs to their corresponding labels. These parameters are typically optimized using backpropagation, which updates them based on the derivatives of a loss function that quantifies the error between the network's predictions and the true targets.

In a DQN, the agent generates its own learning data by interacting with the environment: choosing an action and observing the resulting next state and reward. In this framework, the environmental states serve as the training inputs, and the actions serve as the corresponding labels. At each step $i$, the input nodes of the DQN correspond to the elements of the current state vector $s_i$, while the output nodes represent the estimated action-values $Q(s_i,a)$
 for each action $a$ in the set of allowable actions $\mathcal{A}$. (For simplicity, we assume here that the set of available actions is the same for all states, i.e., $\mathcal{A}(s)=\mathcal{A}$ for all $s\in\mathcal{S}$. Techniques such as action masking~\cite{Huang_CloserLookInvalid_2022, Hou_ExploringUseInvalid_2023, Wang_LearningStateSpecificAction_2024} allow handling state-dependent action sets in deep RL, but these are beyond the scope of this summary.)

After selecting an action $a_i \in \mathcal{A}$, the agent stores the transition $(s_i, a_i, r_{i+1}, s_{i+1})$  in a memory bank of past experiences $\mathcal{D}$, called the \textit{replay buffer}. The network parameters are then updated via backpropagation, using a loss function defined as the average temporal-difference error over a batch of samples drawn from the replay buffer.

Since a neural network is used to approximate the action-value function, we denote the Q-values as $Q(s,a;\bm{\theta})$, where $\bm{\theta}$ collectively represents the network parameters (weights and biases). Recall that traditional Q-learning uses a behavior policy $b$ to update a target policy $\pi$ toward the optimal policy $\pi_*$. In DQN, the $\varepsilon$-greedy behavior policy is based on a network known as the \textit{online} network with parameters $\bm{\theta}$, while the greedy target policy is represented by a separate target network with parameters $\bm{\theta}^-$. The notation $\bm{\theta}^-$ emphasizes that the target network is a delayed copy of the online network parameters $\bm{\theta}$, which stabilizes training.

At each step, the agent selects actions according to an $\varepsilon$-greedy behavior policy, where the greedy choice is determined by the online network’s Q-value estimates:
\begin{equation}
    b^{\rm DQN}(a|s) = 
    \begin{cases}
        1 - \varepsilon + \frac{\varepsilon}{|\mathcal{A}|}, &\text{ if } a=\argmax_{a'\in\mathcal{A}} Q(s,a';\bm{\theta}) \\
        \frac{\varepsilon}{|\mathcal{A}|}, &\text{ otherwise.}
    \end{cases}
    \label{eq:Qlearning_behavior_policy_DQN}
\end{equation}
The loss at the $i$-th step is defined as
\begin{equation}
    L(\bm{\theta}_i, \bm{\theta}_i^-) = \mathbb{E}_{(s,a,r,s') \sim \mathcal{U}(D)} \left[ (Y(r,s'; \bm{\theta}_i^-) - Q(s,a;\bm{\theta}_i))^2 \right],
    \label{eq:DQN_loss}
\end{equation}
where the expectation value is taken over experience tuples $(s,a,r,s')$ sampled uniformly from the replay buffer~$\mathcal{D}$, and the target is defined as
\begin{equation}
    Y(r,s';\bm{\theta}_i^-) = r + \max_{a'\in\mathcal{A}}\gamma Q(s', a';\bm{\theta}_i^-)\,.
    \label{eq:DQN_target}
\end{equation}
Computing the expectation over the entire replay buffer~$\mathcal{D}$ at every learning step would be computationally expensive. Instead, the loss is approximated using a randomly selected subset of experiences, called a \textit{minibatch}, which allows efficient and stable updates of the network parameters.

In traditional Q-learning, Q-values stored in a lookup table are updated directly using Eq.~(\ref{eq:Qlearning_update_rule}). In DQN, the online network parameters are updated at each step $i$ of an episode via backpropagation:
\begin{equation}
\bm{\theta}_i\leftarrow\bm{\theta}_i-\alpha\nabla_{
\bm{\theta}_i}L(\bm{\theta}_i,\bm{\theta}_i^-)\,,
\label{eq:qvalue_network_update}
\end{equation}
where $\alpha$ is the learning rate. The target network parameters in the original DQN algorithm are updated by copying the online network parameters every $C$ steps~\cite{Mnih_HumanlevelControlDeep_2015}. In our implementation, we perform a soft update of the target network at each step:
\begin{equation}
    \bm{\theta}_i^- \leftarrow \tau \bm{\theta}_i^- + (1-\tau) \bm{\theta}_i ,
    \label{eq:target_network_update}
\end{equation}
where $\tau\in(0,1)$ is the \textit{target update rate}. This approach allows the target network to gradually track the online network, improving training stability.

\subsection{Double Deep Q-Learning}
While DQN performs well in many settings, both Q-learning and DQN are susceptible to maximization bias~\cite{Sutton_ReinforcementLearningIntroduction_2018, vanHasselt_DeepReinforcementLearning_2016}. In both methods, the targets, Eqs.~(\ref{eq:Qlearning_target}) and (\ref{eq:DQN_target}), approximate the optimal action-value function $q_*$ by taking a maximum over the estimated values $Q$. Because these estimates contain noise, the maximization step tends to select actions whose values are accidentally inflated. As a result, each update pushes $Q$ toward a target $Y$ that is biased upward. In practice, different states and actions accumulate different amounts of overestimation, and when combined with bootstrapping, this uneven bias can propagate through the value function. Over time, the agent may learn systematically distorted value estimates, causing it to prefer suboptimal states or actions~\cite{vanHasselt_DeepReinforcementLearning_2016}.

To address maximization bias in traditional Q-learning, double learning was introduced~\cite{vanHasselt_DoubleQlearning_2010, vanHasselt_InsightsReinforcementRearning_2011}. Double Q-learning is nearly identical to Q-learning, except it maintains two independent action value tables, $Q^{(A)}$ and $Q^{(B)}$, each updated using different subsets of experience. When updating table $A$, the algorithm does not perform both action selection and action evaluation with the same table. Instead, it first uses table $A$ to select the greedy action, 
\begin{equation}
    a_\mathrm{best}=\argmax_a Q^{(A)}(s_{i+1},a)\,,
\end{equation}
and then uses table $B$ to evaluate that action, forming the target
\begin{equation}
   Y^{(A)}(r_{i+1},s_{i+1}) = r_{i+1} + \gamma Q^{(B)}(s_{i+1},a_\mathrm{best}) .
\end{equation}
Updates to table $B$ are defined symmetrically, with the roles of $A$ and $B$  reversed. Because an action that appears overly valuable in one table is unlikely to be overestimated by the same amount in the other, separating action selection from action evaluation substantially reduces maximization bias.

Extending double learning to deep Q-networks is conceptually straightforward. A Double Deep Q-Network (DDQN) is identical to DQN except that it modifies the target used in the loss function, Eq.~(\ref{eq:DQN_loss}). Instead of using the maximum over target-network values as in Eq.~(\ref{eq:DQN_target}), DDQN decouples action selection from action evaluation:
\begin{equation}
    Y(r,s';\bm{\theta}_i,\bm{\theta}_i^-) = r + \gamma Q(s', \argmax_{a' \in \mathcal{A}}Q(s',a';\bm{\theta}_i);\bm{\theta}_i^-)\,.
    \label{eq:DDQN_target}
\end{equation}
Here, the online network with parameters $\bm{\theta}_i$ selects the greedy action, while the target network with parameters $\bm{\theta}_i^-$ evaluates that action. This mirrors the spirit of double learning, which separates the maximization step across two estimators to reduce overestimation bias.

Unlike traditional double Q-learning, DDQN does not update its two groups of Q-values using disjoint subsets of experiences. Only the online network parameters $\bm{\theta}$ are updated by gradient descent. The target network parameters $\bm{\theta}^-$ are updated either by periodic copying (as in the original DDQN formulation) or by a soft-update rule such as Eq.~(\ref{eq:target_network_update}) (as in our implementation). At first glance, this update mechanism appears inconsistent with the double learning requirement that the two estimators be trained on independent data. However, when the target network is updated infrequently (large update period $C$) or slowly (small soft-update rate $\tau$), the online and target networks remain decorrelated enough for DDQN to capture the benefits of double learning. In practice, this approach significantly mitigates maximization bias while adding minimal computational overhead~\cite{vanHasselt_DeepReinforcementLearning_2016}.

\section{\label{app:our_choice_algo} Our Choice of Algorithm}

Any optimization protocol intended for live, closed-loop learning must meet several demanding requirements: it must avoid getting trapped in local minima, learn efficiently from limited data, operate without explicit knowledge of the underlying system dynamics, and be capable of solving Markov decision processes. In this section, we define off-policy and model-free reinforcement learning and explain how RL algorithms with these properties naturally satisfy the constraints required for real-time optimization. We then compare reinforcement learning in discrete and continuous action spaces, and we argue that deep, discrete RL is particularly well suited for the problem at hand.

In some RL algorithms, learning proceeds by updating the same policy that is used to generate actions. This is known as \textit{on-policy} reinforcement learning: the behavior policy and the policy being improved are identical, and updates are based solely on the most recent state transition. On-policy methods typically rely on exploration strategies such as the $\varepsilon$-greedy policy to escape local minima (see Appendix~\ref{app:Qlearning} for definitions). However, because they learn only from the current transition, they tend to be less sample-efficient. In \textit{off-policy} reinforcement learning, the behavior policy, the one that selects actions, can differ from the target policy being optimized. For example, the target policy might be purely greedy, while the behavior policy is $\varepsilon$-greedy to promote exploration. Transitions generated by the exploratory behavior policy are then used to update the target policy, allowing it to converge toward the optimal policy even if the behavior policy is suboptimal or changing over time.

This separation between behavior and target policies leads to substantially greater sample efficiency. The agent can learn from data generated by any behavior policy, including older versions of itself or even external demonstrators. When combined with a replay buffer, off-policy methods gain an additional advantage: stored transitions can be reused many times to refine value estimates~\cite{Mnih_HumanlevelControlDeep_2015}. Replay buffers also improve stability in deep RL by decorrelating consecutive samples and reducing variance in gradient updates, thereby mitigating harmful feedback loops that can otherwise cause oscillations, convergence to poor local minima, or catastrophic divergence \cite{Mnih_HumanlevelControlDeep_2015}.

Moreover, while model-based RL algorithms assume access to (or learn) an explicit model of the environment's state–transition and reward dynamics to facilitate long-term planning, model-free algorithms make no such assumption. They treat the environment as a black box and learn directly from observed transitions. This allows model-free agents to acquire policies that implicitly capture the full physical dynamics, no matter how complex, without relying on approximate or potentially incorrect models. Such flexibility makes model-free methods well-suited for live optimization of DD sequences in experimental platforms, where the underlying physics may be intricate, imperfectly characterized, or too costly to model and tune explicitly.

Furthermore, RL algorithms with continuous action spaces are generally less sample-efficient than their discrete counterparts~\cite{Recht_TourReinforcementLearning_2019, Wallace_ContinuousTimeReinforcementLearning_2024}, making them less suitable for our setting. This motivated our use of the generators of Thompson’s group $F$ to define a discrete action set, allowing us to leverage the superior sample efficiency of discrete RL. Although our action set is discrete, the state space remains continuous. One could discretize the state space, but doing so would make a table-based representation of the policy or value function intractable for any reasonably fine discretization. Instead, because neural networks are powerful universal function approximators~\cite{Cybenko_ApproximationSuperpositionsSigmoidal_1989, Hornik_MultilayerFeedforwardNetworks_1989, Leshno_MultilayerFeedforwardNetworks_1993}, we employ \textit{deep} reinforcement learning, which represents the relevant functions with neural networks rather than lookup tables.

\section{\label{app:our_agent_and_environment}Our Agent and Environment}

\subsection{States and Rewards}
The environmental state observed by the agent is the ordered sequence of pulse times $(t_1, \ldots,t_N)$. At the start of each episode, the agent is initialized with some starting pulse sequence $(t_1, \ldots,t_N)_\mathrm{initial}$, which has an associated fidelity $p_\mathrm{initial}$ from Eq.~(\ref{eq:avg_prob_of_zero}).

As discussed in Section \ref{sec:intro}, our primary objective is to minimize the wall-clock time required to find an optimal solution. The most time-consuming component of the learning loop in experiment is the evaluation of the reward, since it will depend on experimentally measured estimates of the fidelity, Eq.~(\ref{eq:avg_prob_of_zero}). Consequently, it is crucial to minimize the number of reward evaluations performed during training.

\begin{figure}
    \centering
    \includegraphics[width=\linewidth]{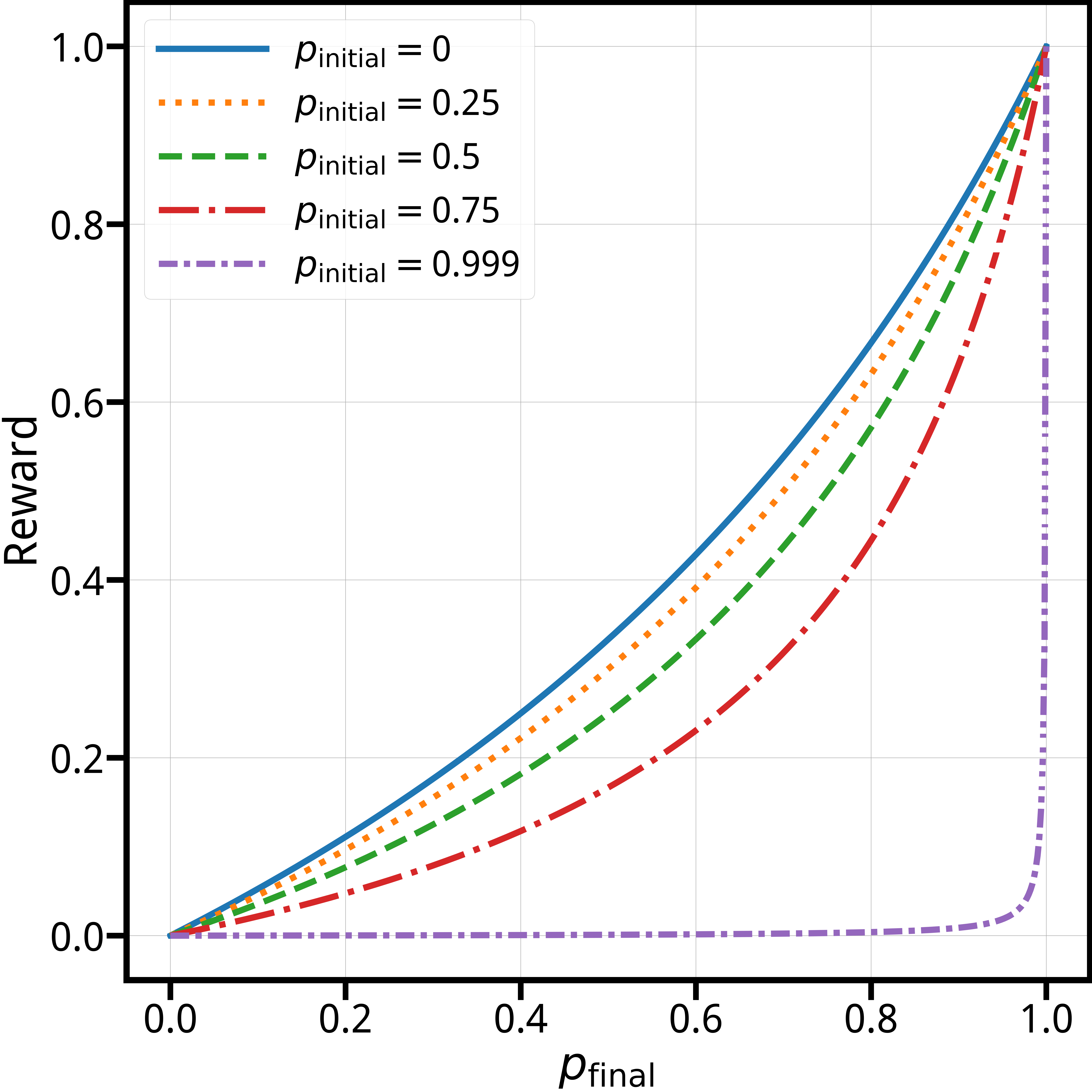}
    \caption{Reward function, Eq.~(\ref{eq:reward}), plotted as a function of the final pulse sequence’s fidelity $p_\mathrm{final}$, shown for different initial pulse sequence fidelities $p_\mathrm{initial}$.}
    \label{fig:reward}
\end{figure}

In our setting, each learning episode terminates after a fixed number of steps $M$. Only at the end of the episode does the agent receive a reward, defined as
\begin{equation}
    r(p_\mathrm{final}) = \frac{(1-p_\mathrm{initial})p_\mathrm{final}}{(1 - p_\mathrm{initial}) + (1-p_\mathrm{final})},
    \label{eq:reward}
\end{equation}
where $p_\mathrm{final}$ is the fidelity corresponding to the final pulse sequence. We choose $M=32$ to strike a balance between providing the agent enough room to explore in the total search space and keeping the problem sufficiently learnable. Figure \ref{fig:reward} shows the reward function as a function of $p_\mathrm{final}$ for several values of  $p_\mathrm{initial}$. Although the reward evaluates to zero when $p_\mathrm{initial}=1$, this poses no practical issue: if the initial pulse sequence already achieves unit fidelity, no dynamical decoupling is needed in the first place.

There are several motivations for defining the reward in this way. First, the reward is normalized to the interval $[0,1]$, and bounded, normalized rewards are known to improve the stability of learning algorithms~\cite{Sutton_ReinforcementLearningIntroduction_2018}. Second, the reward function explicitly favors pulse sequences that outperform the initial sequence, ensuring that the agent is always encouraged to search for improvements. Third, if the same reward function were used for every value of $p_\mathrm{initial}$, then an agent starting with a fidelity already close to one might receive a reward so high that it becomes effectively satisfied with its initial performance, even when substantial relative improvements are still achievable. Because the noise spectrum could, in principle, make the achievable fidelity arbitrarily close to unity, something we cannot know \textit{a priori}, the reward must remain sensitive to small improvements when $p_\mathrm{initial}$ is already large. For this reason, we design the reward function so that its derivative with respect to $p_\mathrm{final}$ near $p_\mathrm{final}=1$ increases as $p_\mathrm{initial}$ approaches 1. This ensures that even modest gains in fidelity are strongly incentivized when the starting performance is already high.

\subsection{Actions}
Our goal is for the agent to transform the initial sequence of pulse times into one that minimizes decoherence. However, since we are using a discrete action space, it is not immediately clear how these actions should be defined.

These efficiency goals impose constraints on our action space that are not trivial to satisfy simultaneously. First, the action set $\mathcal{A}$ should be discrete and as small as possible. A large action set slows training, as the agent must sample more actions to discover optimal solutions, exacerbating the so-called ``curse of dimensionality" in RL, an issue that is even more pronounced with continuous action spaces. Second, the maximum number of steps per episode should be minimized. If rewards are received only after many steps, it becomes difficult for the agent to identify which actions contributed positively or negatively to performance. At the same time, to explore the majority of the environmental state space within each episode, the agent’s actions must be capable of inducing meaningful state changes, while avoiding overly abrupt state transitions that hinder learning.

We therefore define the agent's actions as functions that map the current sequence of pulse times to a new sequence by transforming the entire time interval $[0,T]$ over which they are defined. If each episode consists of $M$ steps, the final pulse sequence $s_M$ is given by a composition of $M$ actions:
\begin{equation}
    s_M = a_{M-1} \circ a_{M-2} \circ \ldots \circ a_1 \circ  a_0(s_0) ,
\end{equation}
where $s_0 = (t_1, \ldots, t_N)_{\mathrm{initial}}$ is the pulse sequence at the start of each episode, and each $a_j$ is selected from a finite action set $\mathcal{A}$. As described in Section \ref{sec:thompsons_group_f}, the full action set is
\begin{equation}
    \mathcal{A} = \{x_0, x_0^{-1}, x_1,x_1^{-1}, \mathrm{id}\}\,,
\end{equation}
where $x_0$ and $x_1$ are the generators of Thompson's group $F$, $x_0^{-1}$ and $x_1^{-1}$ are their inverses, and $\mathrm{id}$ is the identity function on $[0,T]$. Each of the functions $x_0$, $x_1$, $x_0^{-1}$, and $x_1^{-1}$ is appropriately rescaled to act on $[0,T]$ rather than~$[0,1]$.

\subsection{Epsilon Scheduling}
Modern Q-learning algorithms typically do not keep the exploration probability $\varepsilon$ constant across all episodes. Instead, $\varepsilon$ is often adjusted as a function of the number of episodes completed or based on the agent's performance. In our algorithm, we employ a combination of both approaches.

Let the initial reward corresponding to the fidelity $p_\mathrm{initial}$ of the initial pulse sequence be $r_\mathrm{initial} = r(p_\mathrm{initial})$. We initialize both a threshold reward $r_\mathrm{threshold} = r_\mathrm{initial}$ and the maximum reward observed so far, $r_\mathrm{max} = r_\mathrm{initial}$. After each episode, we update $r_\mathrm{max}$ if a higher reward is achieved. Every 50 episodes, if $r_\mathrm{max} > r_\mathrm{threshold}$, we update:
\begin{eqnarray}
    r_\mathrm{threshold} &\leftarrow&r_\mathrm{max} \\
    \varepsilon &\leftarrow& 1 - \frac{r_\mathrm{max} - r_\mathrm{initial}}{1 - r_\mathrm{initial}}\,.
\end{eqnarray}
Since the maximum possible reward is 1, this scheme decreases $\varepsilon$ as $r_\mathrm{max}$ approaches 1, gradually reducing exploration as the agent finds better solutions.

\subsection{Network Topology and Hyperparameters}
\label{app:hyperparameters}
The hyperparameters used to train our agent are summarized in Table~\ref{table:hyperparameters}. Our neural network consists of an input layer, two hidden layers, and an output layer, all employing the ReLU (rectified linear unit) activation function~\cite{Glorot_DeepSparseRectifier_2011, Nair_RectifiedLinearUnits_2010}. The input layer has size $N$, with one neuron corresponding to each pulse time. Each hidden layer contains 32~neurons. The output layer has size~5, with one neuron representing each possible action.

Neuron weights are initialized using uniform Xavier initialization~\cite{Glorot_UnderstandingDifficultyTraining_2010} with an additional gain factor of 0.5. Specifically, the weights are sampled from the uniform distribution $\mathcal{U}(-a, a)$, where
\begin{equation}
    a=0.5\sqrt{\frac6{n_\mathrm{input} + n_\mathrm{output}}} \, .
\end{equation}
Here, $n_\mathrm{input}$ and $n_\mathrm{output}$ denote the number of neurons in the input and output layers, respectively. Biases are sampled from $\mathcal{U}(0,1)$. Stochastic gradient descent is performed using the Adam optimizer~\cite{Kingma_AdamMethodStochastic_2015}.

The replay buffer used has a maximum capacity of 100,000 experiences. When the buffer is full, adding a new experience results in discarding the oldest entry. We define a minimum minibatch size of 25; if the buffer contains fewer than 25 experiences, the agent does not perform neural network backpropagation.

\begin{center}
\renewcommand{\arraystretch}{1.3}
\begin{table}[ht!]
\begin{tabular}{lr}
\hline \hline \\[-10pt]
Hyperparameters & Values \\
\hline \\
Discount factor $\gamma$ & 0.99   \\
Target update rate $\tau$   & 0.01   \\
Learning rate $\alpha$ & 0.0005 \\
Hidden size & 32 \\
Episodes & 5,000 \\
Steps per epsisode & 32 \\
Max buffer size & 100,000 \\
Minibatch size & 128 \\
Minimum minibatch size & 25 \\
Optimizer & Adam \\
Weight initialization & Uniform Xavier \\
Bias initialization & Uniform \\
Activation function & ReLU \\[4pt]
\hline \hline \\
\end{tabular}
\caption{Hyperparameters values for the DDQN agent.}
\label{table:hyperparameters}
\end{table}
\end{center}

\newpage

\nocite{*}

\bibliography{main}

\end{document}